\documentclass[onecolumn,apj,numberedappendix]{emulateapj}
\usepackage{amsfonts}
\usepackage{amsmath}
\usepackage{amssymb}
\usepackage{amsthm}
\usepackage[american]{babel}

\def \TCMB{T_{\rm{CMB}}}
\def \TGAS{T_{\rm{gas}}}
\def \BTGAS{\bar{T}_{\rm{gas}}}
\def \TSPIN{T_{\rm{s}}}
\def \TSTAR{T_*}
\def \TAU{\tau_{\rm{21}}}
\def \TBRIGHT{T_{\rm{21}}}
\def \GUNO{\mathcal G_1}
\def \GDUE{\mathcal G_2}
\def \Tb{T_{\rm b}}
\def \kb{k_{\rm B}}
\def \YA{\bar{Y}_{\alpha}}
\def\spose#1{\hbox to 0pt{#1\hss}}
\def \lta{\mathrel{\spose{\lower 3pt\hbox{$\mathchar"218$}}
     \raise 2.0pt\hbox{$\mathchar"13C$}}}
\def\gta{\mathrel{\spose{\lower 3pt\hbox{$\mathchar"218$}}
     \raise 2.0pt\hbox{$\mathchar"13E$}}}
\def \lya{Ly$\alpha\ $}
\def\be{\begin{equation}}
\def\ee{\end{equation}}

\lefthead{Pillepich, Porciani \& Matarrese}
\righthead{The bispectrum of 21-cm fluctuations}


\begin{document}
\submitted{}

\title{The bispectrum of redshifted 21-cm fluctuations from the dark ages}
\author{Annalisa Pillepich\altaffilmark{1}, Cristiano Porciani\altaffilmark{1} and Sabino Matarrese\altaffilmark{2}}
\altaffiltext{1}{Institute for Astronomy, ETH Z\"urich, 8093 Z\"urich, 
Switzerland}
\altaffiltext{2}{Dipartimento di Fisica ``G. Galilei'', Universit\`a di 
Padova, and INFN Sezione di Padova, via Marzolo 8, I-35131 Padova, Italy}



\begin{abstract}
Brightness-temperature fluctuations in the redshifted 21-cm background
from the cosmic dark ages are generated by irregularities in 
the gas-density distribution and can then be used to determine the statistical 
properties of density fluctuations in the early Universe. 
We first derive the most general expansion of brightness-temperature fluctuations
up to second order in terms of all the possible sources of spatial fluctuations.
We then focus on the three-point statistics and compute the angular bispectrum of 
brightness-temperature fluctuations generated prior to the epoch of hydrogen 
reionization.  For simplicity, we neglect redshift-space distortions.
We find that low-frequency radio 
experiments with arcmin angular resolution can easily detect
non-Gaussianity produced by non-linear gravity with high signal-to-noise ratio.
The bispectrum  thus provides a unique test of 
the gravitational instability scenario for structure formation, 
and can be used to measure the cosmological parameters.
Detecting the signature of primordial non-Gaussianity 
produced during or right after
an inflationary period is more challenging but still possible.
An ideal experiment limited by cosmic variance only and 
with an angular resolution of a few arcsec
has the potential to detect primordial non-Gaussianity with 
a non-linearity parameter of $f_{\rm NL}\sim 1$.
Additional sources of error as
weak lensing and an imperfect foreground subtraction could
severely hamper the detection of primordial non-Gaussianity
which will benefit from the use of 
optimal estimators combined with tomographic techniques.
\end{abstract}


\section{Introduction}

During the dark ages (the time between recombination and the formation of the first stars), 
the cosmic microwave background (CMB) is coupled to atoms of neutral hydrogen through spin-flip
21-cm transitions. 
Due to the resonant nature of the interaction,
neutral hydrogen at redshift $z$ imprints a signature at a wavelength
of $21.12\,(1+z)$ cm in the CMB.
The brightness temperature of the CMB at radio wavelengths
thus probes the three-dimensional neutral hydrogen distribution at $30<z<100$.
This accurately traces dark-matter inhomogeneities down to the Jeans length
($\sim 10$ comoving pc corresponding to angular separations of $\sim 10^{-2}$ arcsec).
On smaller scales, the finite pressure of the gas keeps the baryons
uniformly distributed.
The expected power-spectrum of fluctuations in the 21-cm background from the dark ages
has been recently calculated (Loeb \& Zaldarriaga 2004; Bharadwaj \& Ali 2004).
In this paper, we focus on the three-point statistics. In particular, we compute 
the bispectrum of 21-cm maps. This is obtained by expanding the brightness 
temperature of the cosmic background 
up to second order in the underlying density fluctuations.
Related work has been presented by Cooray (2005) who discussed 
the large-scale non-Gaussianity generated during the era of reionization and
by Saiyad, Bharadwaj \& Pandey (2006) who focussed on the post-reionization
era ($z<6$). 

Measuring the bispectrum of fluctuations in the 21-cm background will ultimately 
allow to quantify the degree of non-Gaussianity (NG) of the dark matter 
density field at high-redshift. 
Such a non-Gaussianity has two possible origins: non-linearity due to 
the usual Newtonian gravitational instability 
and non-Gaussianity which is intrinsic in the mechanism generating 
the primordial seeds, i.e. arising during or immediately after inflation, 
as well as post-Newtonian terms inherent in the 
solution of Einstein's equations to second order 
(Bartolo, Matarrese \& Riotto 2005). 
The latter contributions to non-Gaussianity are usually expressed 
through a {\it non-linearity parameter} $f_{\rm NL}$ which measures 
the strength of quadratic terms in the {\it primordial} gravitational 
potential (e.g. Salopek \& Bond 1990, 1991; Gangui et al. 1994;
Verde et al. 2000a). 
The most stringent limits on $f_{\rm NL}$ are based on the analysis of 
CMB anisotropies from three-year WMAP data 
(Spergel et al. 2006) and give $-54 < f_{\rm NL} < 114$ (3$\sigma$ confidence level). 
The high-resolution observations of CMB anisotropies by the 
{\it Planck} satellite should reduce the 3$\sigma$ detection threshold  
to $|f_{\rm NL}| \sim 10$. 
Searches for primordial NG based on the large-scale structure 
of the Universe where the relevant observables probe the dark matter 
density field (e.g. three-point statistics of the 
galaxy distribution) are plagued with two problems.
First, the density field is related to the gravitational potential through the Poisson equation
so that its Fourier modes weigh the NG contributions with extra $k^{-2}$ terms and only the largest 
scales may keep memory of primordial NG. Moreover, the ratio of Newtonian non-linear 
terms to intrinsical NG terms roughly scales like $(1+z)^{-1}$, so 
that NG signatures in the local Universe are easily masked by the effects of
gravitational instability. In this sense, the 21-cm background at large 
redhifts appears as an extremely promising dataset to search for primordial 
NG able to provide constraints on $f_{\rm NL}$ which can be  
complementary and possibly competitive with those based on CMB anisotropies.

Low-frequency radio observations with high angular resolution are limited by 
the Earth's ionosphere. This layer of the atmosphere
is opaque to electromagnetic waves whose frequency lies below the local plasma frequency
which corresponds to $\sim 15$ MHz during day-time and near sunspot maximum
and to $\sim 10$ MHz at night near sunspot minimum.
These figures are somewhat lower (down to $\sim 2$ MHz) at preferred sites located near the magnetic 
poles but, in general, the propagation properties of radio waves with
frequencies $\nu<30$ MHz show extreme variations due to ionospheric scintillation 
(e.g. Kassim et al. 1993).
Therefore, it is unlikely that 
ground-based experiments with high angular resolution ($<1$ arcmin) will ever measure the properties
of the 21-cm background at a redshift $z\gg50$.
Space observatories are needed to overcome these limits.

The layout of the paper is as follows.
In \S\ref{21} we summarize the physics of 21-cm radiation and perform the expansion
of its brightness temperature to the second perturbative order in all the relevant quantities. 
Non-Gaussian density fields are introduced in \S\ref{nongau}.
The calculation of the angular bispectrum of redshifted 21-cm fluctuations is presented in \S\ref{calc}.
Finally, in \S\ref{concl}, we estimate the detectability of non-Gaussian features (both primordial
and gravity induced) by future experiments and conclude.

We consider a ``concordance'' cosmological model with a present-day
matter-density parameter 
$\Omega_{\rm m}=0.27$ (of which $\Omega_{\rm b}=0.049$
in baryons), 
a cosmological costant contribution of $\Omega_\Lambda=0.73$ and a 
Hubble constant
of $H_0=100\,h\,{\rm km}\,{\rm s}^{-1}\,{\rm Mpc}^{-1}$ with $h=0.71$.
We also assume a Harrison-Zel'dovich primordial power spectrum with spectral index $n=1$ and 
a cold dark matter (CDM) 
transfer function with present-day normalization $\sigma_8=0.9$ 
(where $\sigma_8$ denotes the linear 
rms fluctuation within a sphere with a comoving radius of 8 $h^{-1}$ Mpc).

\section{The 21-cm fluctuations}
\label{21}
\subsection{Kinetic temperature of the IGM}
About $3\times 10^5$ yr after inflation, atomic nuclei and electrons combined to form neutral atoms, and greatly reduced their
coupling with photons. Soon after, when the free-streaming CMB radiation cooled below 3000 K and shifted into the
infrared, the Universe became dark.
The cosmic dark
ages persisted until the first luminous sources formed inside virialized density perturbations ($z\sim 30$), 
and reionized the opaque IGM. 
The exact timing of this event is still uncertain, even though observational data bracket it between
redshift 20 and 6 (e.g. Loeb \& Barkana 2001; Spergel et al. 2006).

The kinetics of recombination and ionization processes in a rapidly 
expanding Universe leaves a residual degree of ionization even during the dark ages.
At $z\simeq 200$, the mean hydrogen ionization fraction ${\bar x}\simeq 10^{-4}$ while the fraction of singly ionized
Helium is much lower ($\simeq10^{-9}$). These figures slightly decrease with time due to the cosmic
expansion but always remain of the same order of magnitude (e.g. Seager et al. 2000 and references therein).

Compton scattering of CMB photons off free electrons couples 
the kinetic temperature of cosmic gas $\TGAS$ with the 
temperature of the photon background (Weymann 1965)
%
\be
\frac{\partial\TGAS}{\partial t}-\frac{2}{3}\,\frac{\TGAS}{n_{\rm tot}}\,\frac{\partial n_{\rm tot}}{\partial t}
=\frac{8}{3}\, \frac{\sigma_{\rm T}\, U\, n_{\rm e}}{m_{\rm e}\, c\, n_{\rm tot}} \left( \TCMB-\TGAS\right)\;,
\label{evoTgas}
\ee
where $\sigma_{\rm T}$ is the Thomson scattering cross section,  $m_{\rm e}$ the electron mass and $c$ the
speed of light in vacuum. The total number density of particles, $n_{\rm tot}$, appears on the right-hand side 
because collisions
and Coulomb scatterings hold all the different species (electrons, ions and atoms) at the same temperature.
Here $U$ represents the radiation energy density
$U=\int_0^\infty u_\nu\,d\nu$ where $u_\nu=4\pi J_\nu/c$ and
$J_\nu=(1/4\pi)\int_{4\pi} I_\nu\, d\hat{n}$ 
with $I_\nu$ the specific intensity of radiation (energy per unit time, surface, frequency and solid angle) and $d\hat{n}$ the solid-angle element.
In thermal equilibrium, the radiation field has a frequency distribution given
by the Planck function $J_\nu=B_\nu(\TCMB)$ and the total energy is
given by Stefan's law $U=(\pi^2\,k_{\rm B}^4/15\,\hbar^3\,c^3)\,\TCMB^4$
where $k_{\rm B}$ and $\hbar$ denote the Boltzmann and the Planck constants, respectively.
Note that the spectrum of the CMB remains close to a blackbody because the 
heat capacity of radiation is very much
larger than that of matter (i.e. there are vastly more photons than baryons).
For a uniform distribution of baryons with physical number density $n_{\rm{tot}} \propto (1+z)^3$,
Compton scattering is efficient down to a redshift $z \sim 150 \,(\Omega_{\rm b}h^2/0.023)^{2/5}$ 
(the decoupling era, e.g. Peebles 1993) and keeps $\TGAS=\TCMB=2.73\, (1+z)$ K. 
Subsequently, the gas expands adiabatically, $\TGAS \propto (1+z)^2$.

\subsection{Spin temperature of neutral hydrogen}
\label{spintempneuh}
Neutral hydrogen interacts with radiation of wavelength $\lambda_{\rm{21}}$= 21.12 cm through 
the resonant transition between the two hyperfine levels of the 1s state.
This corresponds to 
a frequency 
$\nu_{\rm{21}} = 1420.4 $ MHz and a temperature $\TSTAR= 0.068 $ K.
It is convenient to express
the ratio between the number density of hydrogen atoms in the excited state, $n_*$,and 
in the ground state, $n_{\circ}$, in terms of a Boltzmann factor which defines the spin temperature, $\TSPIN$:
\begin{equation}
\label{spintem}
\frac{n_*}{n_\circ} =\frac{g_*}{g_\circ}\,  \exp{\left(-\frac{\TSTAR}{\TSPIN}\right)},
\end{equation}
where $g_*=3$ and $g_\circ=1$ are the spin-degeneracy factors of the excited and fundamental levels 
in the 1s state.
The total density of neutral hydrogen atoms is  $n_{\rm{HI}}=n_\circ+n_*$.
During the dark ages, the time evolution of $n_\circ$ is the result of the combined 
effect of atomic collisions and radiative interactions with the photon background.
Approximating the 21-cm transition as infinitely sharp, one gets
\begin{equation} 
\left ( \frac{\partial}{\partial t} + 3\,\frac{\dot{a}}{a}  \right) n_\circ ~ = ~ -n_\circ\,(C_{\rm{01}}+ B_{\rm{01}}\,4\pi\,J_{\nu_{\rm{21}}}) + n_*\,(C_{\rm{10}}+A_{\rm{10}}+ B_{\rm{10}}\,4\pi\,J_{\nu_{\rm{21}}})\;,
\end{equation}
where
$a(t)=(1+z)^{-1}$ is the cosmic scale factor, $A_{10}=2.85\times 10^{-15}$ s$^{-1}$ is
the spontaneous decay rate of the hyperfine transition of atomic hydrogen, 
$C_{\rm{01}}$ and $C_{\rm{10}}$ are the collisional excitation and 
de-excitation rates, while $B_{\rm{01}}$ and $B_{\rm{10}}$ are the Einstein rate coefficients.
At equilibrium, detailed balancing of collisional and radiative processes gives
\be
\frac{C_{01}}{C_{10}}=\frac{g_*}{g_\circ}\,  \exp{\left(-\frac{\TSTAR}{\TGAS}\right)}\;,
 \ \ \ \ \ \ \
\frac{A_{10}}{B_{01}}=\frac{g_\circ}{g_*} \,4\pi\,F_{\nu_{\rm{21}}}\;, \ \ \ \ \ \ \   
\frac{B_{10}}{B_{01}}=\frac{g_\circ}{g_*}\;,
\ee
with $F_\nu=4\pi\,\hbar\,\nu^3/c^2$.
When $\TGAS, \TCMB$ and $\TSPIN$ are much larger than $\TSTAR$, all the Boltzmann and the Bose-Einstein factors
can be linearized, so as to obtain:
\be
\label{evost}
\frac{1}{1+(g_*/g_\circ)}\,\frac{d}{dt} \left(\frac{\TSTAR}{\TSPIN} \right) =
C_{10}\left(\frac{\TSTAR}{\TGAS}-\frac{\TSTAR}{\TSPIN} \right) +A_{10}
\left(1-\frac{\TCMB}{\TSPIN} \right)\;,
\ee
where the collisional de-excitation rate, $C_{10}=(4/3)~\kappa_{10} ~n_{\rm{HI}}$,
with $\kappa_{\rm{10}}$ a function of $\TGAS$ tabulated by Zygelman (2005). 
%
At mean density, the characteristic timescales for reaching an equilibrium state are
$\TSTAR/(A_{10}\,\TCMB)\simeq  5.4 \times 10^3\, [(1+z)/51]^{-1}$ yr for radiative processes
and $C_{10}^{-1}\simeq  1.2 \times 10^4\,(1+\delta)^{-1}[(1+z)/51]^{-5}$ yr for atomic collisions
(assuming that $\kappa_{10}\propto \TGAS$ which is accurate for $z\sim 50$ where $\TGAS \sim 50$ K).
Since these timescales are much shorter than the Hubble time, for cosmological studies one is only interested in the  
steady state solution of equation (\ref{evost})
\be
\TSPIN = \frac{\TCMB+Y_{\rm{c}} ~\TGAS }{1+Y_{\rm{c}}}\;, \ \ \ \ \ {\rm with} \ \ \ \ \
Y_{\rm{c}} = \frac{C_{\rm{10}}~ 
\TSTAR}{A_{\rm{10}}~\TGAS}\;. 
\ee
In other words, atomic collisions drive $\TSPIN$ towards $\TGAS$ while 21-cm transitions drive it towards $\TCMB$.
Which process dominates depends on the local gas density and collisions become less and less effective
with time. For regions at mean density, $\TSPIN$ departs from $\TGAS$ at $z\sim 100$ and approaches
$\TCMB$ at $z\sim30$. On the other hand, at $z\sim 50$ collisions are efficient only within regions with
overdensities such that $1+\delta > 2.2 \,[(1+z)/51]^{-4}$.

This picture is modified by the appearance of the first luminous objects at $z\lta 30$.
At this epoch, bubbles of ionized material start appearing around the sources of light while
neutral gas is heated by collisions with fast electrons produced by penetrating photons (X-rays and UV radiation).
The detailed evolution of both these phenomena depends on uncertain astrophysical details 
about the formation of massive stars,  X-ray binaries and accreting black holes.
Anyway,
the presence of \lya photons provides an additional coupling between $\TSPIN$ and $\TGAS$
through the so called Wouthuysen-Field effect (Wouthuysen 1952; Field 1958).
According to this mechanism, the populations of the singlet and triplet 1s states can be mixed up
by the absorption of \lya photons followed by a radiative decay.
The efficiency of the effect is determined by the frequency dependence and the intensity
of the background radiation field $J_\nu$ near \lya.
This is conveniently expressed in terms of the coupling coefficient
$Y_\alpha$ (the \lya pumping efficiency) and the color temperature $T_\alpha$ defined as
\begin{equation}
Y_\alpha = \frac{P_{10}}{A_{10}} \frac{T_\ast}{T_\alpha}\;,
\ \ \ \ \ \ \ \frac{1}{k_{\rm B}\,T_\alpha}=-\frac{\partial \log N_\nu}{\partial h\nu}
\end{equation}
where $ N_\nu=J_\nu/F_\nu$ gives the photon occupation number at frequency $\nu$ 
and  $P_{10}=(4/27) P_\alpha$ 
is the indirect de-excitation rate of the triplet via absorption of a \lya photon to the $n=2$ level
with $P_\alpha=4\pi\,\int \sigma_\nu\,(J_\nu/2\pi\hbar\nu)\,d\nu$ the total \lya scattering
rate and $\sigma_\nu$ the cross section for \lya scattering 
(Madau et al. 1997).
In steady state, 
the spin temperature of neutral hydrogen is a weighted average of $\TGAS$, $\TCMB$ and $T_\alpha$
\begin{equation}
\label{SPINTEMP}
\TSPIN = \frac{\TCMB+Y_{\rm{c}} ~\TGAS + Y_{\alpha}~ T_{\alpha}}{1+Y_{\rm{c}}+Y_{\alpha}}\;.
\end{equation}
There exists then a critical value of $P_\alpha=27\,A_{10}\,\TCMB/4\,\TSTAR\simeq
8.5\times 10^{-12}\,[(1+z)/11]$ s$^{-1}$
above which $\TSPIN\simeq T_\alpha$. This roughly corresponds to $\sim 1$ \lya photon per baryon.
Note that, in general, background photons emitted as continuum, non-ionizing UV radiation and redshifted into 
the Ly$\alpha$ resonance represents 
the dominant source of coupling for $\TSPIN$ 
with respect to line photons locally emitted via recombinations and collisional excitations (Madau et al. 1997).
%
Moreover,
due to the large cross section for resonant scattering and the recoil of the atoms during photon emission,
thermalization of the radiation is very efficient and in virtually all the physically relevant situations $T_\alpha=\TGAS$
(Field 1959b).

\subsection{Brightness temperature of 21-cm fluctuations}
The brightness temperature $\Tb$ of a radiation field is defined as $I_\nu=B_\nu(\Tb)$ and,
in the Rayleigh-Jeans limit ($h\nu/\kb \Tb \ll 1$), $\Tb=c^2\, I_\nu/2\,\nu^2\kb$.
In the absence of energy exchanges between matter and radiation after decoupling, 
the brightness temperature of the CMB would be $\Tb=\TCMB$.
However, interactions between CMB photons and atoms of neutral hydrogen along the line of sight 
modify the brightness temperature of the CMB at radio wavelengths.
The optical depth for the hyperfine transition at $\lambda=\lambda_{21}\,(1+z)$ 
(once again approximated as infinitely sharp) is (Field 1959a)
\begin{equation}
\label{OPTICALDEPTH}
\TAU= \frac{3\, c^3\, \hbar\, A_{\rm{10}}\, n_{\rm{HI}}}{16\, k_{\rm{B}}\, \nu_{\rm{21}}^2\, 
\TSPIN \,\left (\partial v_{\rm{r}}/\partial r_{\rm{phys}} \right)}
\end{equation} 
where the number density of neutral hydrogen, the spin temperature and the radial velocity gradient
(in physical units) 
are both evaluated at a comoving distance $r(z)$ along the line of sight.
This is obtained in the Sobolev approximation to account for line photons redshifted in or out
of the interaction frequency range by the combined action of Hubble expansion and peculiar velocities
(e.g. Sobolev 1960)
%
In the Rayleigh-Jeans limit, 
the radiative transfer equation for the rest-frame brightness temperature of a patch of the sky 
at redshift $z$ gives
\begin{equation}
\Tb= \TCMB ~e^{-\TAU}+\TSPIN~(1-e^{-\TAU})\;.
\end{equation}
The effect of the 21-cm spin-flip transition can then be measured
as an excess or a decrement of the brightness temperature of the sky (as observed at the present time) 
at a wavelength $\lambda=\lambda_{21}\,(1+z)$
(Scott \& Rees 1990):
%
\begin{equation}
\label{21BRIGHTNESSTEMP}
\TBRIGHT= \frac{T_{\rm{b}} - \TCMB}{1+z}
\simeq \frac{\TSPIN-\TCMB}{1+z}~ \TAU\;.
\end{equation}
The last equality holds in the optically thin limit which
always applies since, neglecting peculiar velocities,
\be
\TAU\simeq 0.025\, \frac{\TCMB}{\TSPIN}\,
\left(\frac{\Omega_{\rm b}\,h}{0.035}\right)\,
\left(\frac{\Omega_{\rm m}}{0.27}\right)^{-1/2}
\left(\frac{1+z}{51}\right)^{1/2}\ll 1\;.
\ee
The spatial dependence of $\TBRIGHT, \TSPIN, \TCMB$ and $\TAU$ is understood in equation (\ref{21BRIGHTNESSTEMP}) but
all these quantities should be evaluated at a comoving distance $r(z)$ along the line of sight.
Note that intrinsic fluctuations of $\TCMB$ can be fully neglected in this context since
they are of the order of a few $\mu$K while those of $\TBRIGHT$ are roughly a thousand times larger.
Given the thermal history of the gas, four regimes can be distinguished: 
{\it i)} the pre-decoupling era ($z \gta 200$) 
when $\TSPIN=\TCMB$ and no fluctuations are present in the brightness temperature of the 21-cm  background;
{\it ii)} the dark ages ($30 \lta z \lta 200)$ when $\TSPIN \ll \TCMB$ and neutral hydrogen absorbs CMB photons;
{\it iii)} the reionization epoch ($10 \lta z \lta 30$) when $\TSPIN \gg \TCMB$ and neutral hydrogen emits CMB photons;
{\it iv)} the post-reionization era ($z\lta 10$) when the IGM is nearly fully ionized over most of its volume 
and $\TBRIGHT=0$ almost everywhere (with the exception of the highest density regions).

\subsection{Fluctuations of the brightness temperature}
\label{FLUCT}
The brightness temperature of the CMB at $\lambda=\lambda_{21}\,(1+z)$ cm fluctuates with the position on the sky
and redshift reflecting the variation of a number of underlying quantities.
An excess in the density of neutral hydrogen (either determined by an overdensity $\delta$
or by a fluctuation of the ionized fraction $\delta_x$),
\be
n_{\rm{HI}}=\bar{n}_{\rm{H}} ~[1-\bar{x}~(1+\delta_x)]~(1+\delta)\;,
\ee
produces an increment of the optical depth and alters $\TSPIN$ (directly, by increasing the collision rate, and
indirectly, by changing the gas temperature).
Radial peculiar velocities, $v_{\rm pec}$, also modify $\TAU$ through the term
\be
\left( \frac{\partial v_{\rm{r}}}{\partial r_{\rm{phys}}} \right)= 
H(z)+ (1+z)\, \frac{\partial v_{\rm{pec}}}{\partial r}\;,
\ee
where $H(z)$ is the Hubble parameter at redshift $z$.
Finally, during reionization, also fluctuations in the intensity and temperature of the \lya background
become important.

For small fluctuations,
we perform a perturbative expansion of the interesting quantities by writing $\delta_i=\sum_n \delta_i^{(n)}/n!$
where $\delta_i^{(n)}={\cal O}[(\delta_i^{(1)})^n]$ and the subscript $i$ indicates a particular random field
(e.g. temperature, density, etc.). 
In order to compute the first non-vanishing term of the bispectrum we need to expand $\TBRIGHT$
up to $2^{\rm nd}$ order in 
$\delta$, $\delta_x$, $v_{\rm{pec}}$, and $\delta_{\rm{\alpha}}$ (i.e. the fluctuation in the pumping
efficiency of the \lya background, $Y_\alpha$).
Fluctuations of $\TGAS=\BTGAS ~(1+  \delta_T)$
can be expressed in terms of the density perturbations through the following functional expansion in real space:
\be
\label{GUNOGDUERS}
\delta_T(\mathbf{r},z)= \int d^3\mathbf{x_1} \,\, \GUNO(\mathbf{x_1},z) \, \delta(\mathbf{r}+\mathbf{x_1},z)+
\int d^3\mathbf{x_1} \int d^3\mathbf{x_2} \,\, \GDUE(\mathbf{x_1},\mathbf{x_2},z) \, 
\delta(\mathbf{r}+\mathbf{x_1},z) \, \delta(\mathbf{r}+\mathbf{x_2},z)+...
\ee
where the dots denote $n$-point non-local dependencies with $n>2$. 
Both kernels $\GUNO$ and $\GDUE$ vanish before decoupling when Compton scattering keeps 
$\TGAS=\TCMB$.
On the other hand, at late times, when the gas expands adiabatically, all the terms in equation
(\ref{GUNOGDUERS}) are local: $\GUNO(\mathbf{x},z)=(2/3)\,\delta_{\rm D}(\mathbf{x})$ and
$\GDUE(\mathbf{x_1},\mathbf{x_2},z)=-(2/9)\,\delta_{\rm D}(\mathbf{x_1})\,\delta_{\rm D}(\mathbf{x_2})$
where $\delta_{\rm{D}}({\bf x})$ is the Dirac delta function.
The complete redshift evolution between these asymptotic regimes 
is determined by equation (\ref{evoTgas}), that can be rewritten as
\be
\frac{\partial\delta_{\rm{T}}}{\partial t}-
\frac{2}{3}\,\frac{1+\delta_{\rm{T}}}{1+\delta}\,\frac{\partial\delta}{\partial t}
=-C(t)\,
\delta_{\rm{T}};
\label{evoTgas2}
\ee
where,
$C(t)=(8\,\sigma_{\rm T}\, U\, n_{\rm e}\,\TCMB)/(3\,m_{\rm e}\, c\, n_{\rm tot}\,\BTGAS )$.
Solutions for the kernels $\GUNO$ and $\GDUE$ are derived in Appendix \ref{PGT} where it is
shown that $\GUNO(\mathbf{x},z)=g_1(z)\,\delta_{\rm D}(\mathbf{x})$ at all epochs while $\GDUE$ has
a complex spatial dependence.
These results extend the linear analysis by Bharadwaj \& Ali (2004)\nocite{BA2004a} to second order
(our function $g_1$ corresponds to their linear parameter $g$).
Note that these authors use a similar technique to relate fluctuations in $\TSPIN$ and $\delta$
through equation (\ref{evost}).
Here, instead, we perturb the steady state solution given in equation (\ref{SPINTEMP}) which allows us
to explicit the dependence on the underlying variables in analytic form.

\begin{figure}
\plottwo{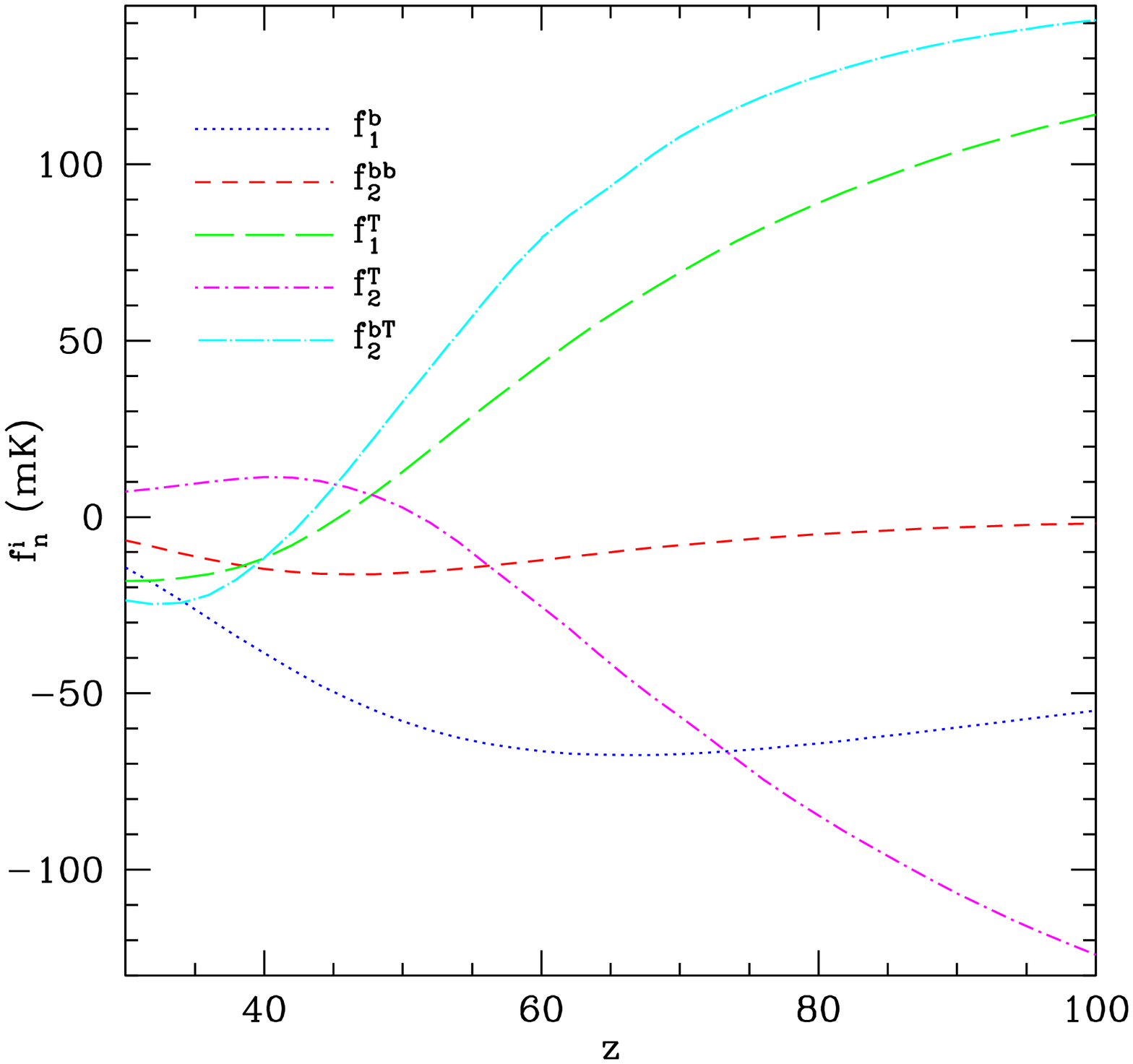}{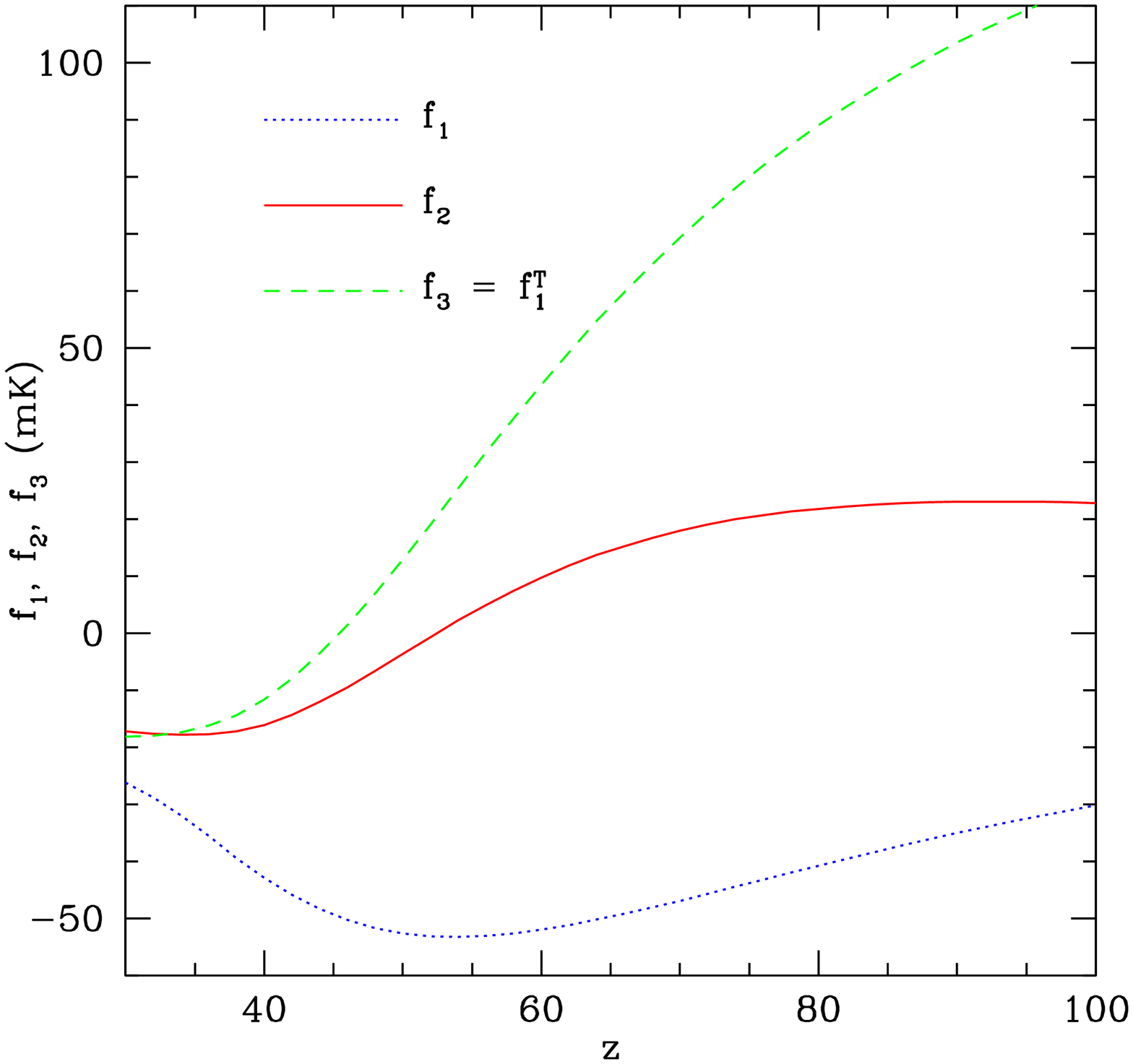}
\caption{\label{F12}Left: Redshift evolution of the coefficients that couple the brightness-temperature fluctuations 
to perturbations in the gas temperature and density (see the main text and both the Appendices for details).
Right: Evolution of the ``effective'' coupling coefficients $f_1$, $f_2$ and $f_3=f^{\rm{T}}_1$  introduced in 
equation (\ref{T21}).
}
\end{figure}

The most general expansion of $\TBRIGHT$ in terms of all the possible sources of spatial fluctuations up to
 second order is reported in Appendix \ref{SOET}.
Here, we focus on the 21-cm signal coming from the dark ages. We therefore neglect fluctuations in the \lya 
background and we consider inhomogeneities in the hydrogen ionization fraction whose contribution is depressed 
by a factor of  $\sim \bar{x}/(1-\bar{x})\simeq 10^{-4}$ with respect to that generated by density fluctuations.
As a first step, we also ignore peculiar velocities, postponing to future work a detailed study of their 
subdominant effects.
In this case, the spatial fluctuations of the brightness temperature can be written as
\begin{eqnarray}
\label{T21}
\Delta\TBRIGHT(\mathbf{r},z) &=& \TBRIGHT(\mathbf{r},z)-\TBRIGHT^{(0)}(z)=\\
\nonumber
&=& f_1(z)\,\left( \delta^{(1)}(\mathbf{r},z)+ \frac{1}{2}\, \delta^{(2)}(\mathbf{r},z)\right)+ f_2(z)~\, \delta^{(1)}(\mathbf{r},z)^2+\\
&+& f_3(z)\,\int d^3\mathbf{x_1} \int d^3\mathbf{x_2} \,\, \GDUE(\mathbf{x_1},\mathbf{x_2},z) \, \delta(\mathbf{r}+\mathbf{x_1},z) \, \delta(\mathbf{r}+\mathbf{x_2},z) \;,
\nonumber
\end{eqnarray}
where
\begin{equation}
\TBRIGHT^{(0)}= f_{\rm{0}}\left( 1- \frac{\TCMB}{\mathcal{S}} \right) (1- \bar{x})
\end{equation}
and, since  $H(z)\simeq H_0\,\Omega_{{\rm m}}^{1/2}\,(1+z)^{3/2}$ for $z\gg 1$,
\be
f_{\rm{0}}= \left( \frac{3 c^3 \hbar A_{\rm{10}}}{16 k_{\rm{B}} \nu_{\rm{21}}} \right) \frac{\bar{n}_{\rm{HI}}}{H_{\rm{0}}} {\left( \frac{1+z}{\Omega_{{\rm m},\circ}} \right)}^{1/2}
\simeq 69.05\ {\rm mK}\ 
\left(\frac{\Omega_{\rm b}\,h}{0.035}\right)\,
\left(\frac{\Omega_{\rm m}}{0.27}\right)^{-1/2}
\left(\frac{1+z}{51}\right)^{1/2}
\;. 
\ee
The functions $f_1$, $f_2$ and $f_3$ are different combinations 
of the coefficients that couple the brightness temperature
with the local baryonic density (superscript b) and the gas temperature (superscript T):
\begin{eqnarray}
\label{F1}
f_1&=&f^{\rm{b}}_1+g_1~f^{\rm{T}}_1;\\
\label{F2}
f_2&=&f^{\rm{bb}}_2+g^2_1~f^{\rm{TT}}_2+g_1~f^{\rm{bT}}_2;\\
\label{F3}
f_3&=&f^{\rm{T}}_1.
\end{eqnarray}
Analytic expressions for these coupling constants are given 
in Appendix \ref{SOET} and their behaviour  as a function of redshift is shown in Figure \ref{F12}.
An overdensity of order unity gives rise to a fluctuation in $\TBRIGHT$ of a few tens of mK and
the expected signal is maximum at $z\simeq 50$. Note 
that $f_2$ and $f_3$ change sign with redshift and multifrequency experiments can be used to isolate
the contributions of the different terms in equation (\ref{T21}).

\section{Non-Gaussian density fields}
\label{nongau}
\subsection{The bispectrum}
From now on it will be convenient to work in Fourier space. We define the Fourier transform of the random field
$X(\mathbf{r})$ as $\tilde X(\mathbf{k}) =\int  d^3 \mathbf{r} ~X(\mathbf{r})\,
\exp\left(- i \mathbf{k}\cdot\mathbf{r}\right)$, where $\mathbf{r}= r \hat{\bf n}$.
For a stationary random field (statistically invariant under rotations and translations),
the power spectrum $P(k)$ is given by the ensemble average
\be
\langle \delta \tilde{X}(\mathbf{k_1})~\delta \tilde{X}(\mathbf{k_2}) \rangle
= (2\pi)^3\, P(k_1) ~\delta_{\rm{D}}(\mathbf{k_1}+\mathbf{k_2})\;.
\ee
Similarly, the bispectrum $ B(\mathbf{k_1},\mathbf{k_2},\mathbf{k_3})$ is defined as
\begin{equation}
\label{3DBISP}
\langle \delta \tilde{X}(\mathbf{k_1})~\delta \tilde{X}(\mathbf{k_2})~\delta \tilde{X}(\mathbf{k_3}) \rangle  = (2\pi)^3 \,B(\mathbf{k_1},\mathbf{k_2},\mathbf{k_3}) ~\delta_{\rm{D}}(\mathbf{k_1}+\mathbf{k_2}+\mathbf{k_3})\;.
\end{equation}
For a Gaussian random field, $B=0$.
\subsection{Non-Gaussianity from gravitational instability}
For the range of scales and redshifts considered here,  
the baryons trace the dark matter, so that we can speak of an unique overdensity
$\delta = \delta^{(1)} + \frac{1}{2}\delta^{(2)}+{\cal O}(\delta^{(3)})$, 
where the Fourier-space expressions for the first and second-order terms read,
respectively
\begin{equation}
\label{1BDFS}
\tilde{\delta}^{(1)}(\mathbf{k},z) = 
D_+(z)\, \tilde{\delta}^{(1)}_{0}(\mathbf{k})
\end{equation}
and  
\begin{equation}
\label{2BDFS}
\frac{1}{2}\,\tilde{\delta}^{(2)}(\mathbf{k},z) = 
\frac{D^2_+(z)}{(2\pi)^3}\, \int d^3\mathbf{q_1}\int d^3\mathbf{q_2}\, 
\,\delta_{\rm{D}}(\mathbf{q_1}+\mathbf{q_2}-\mathbf{k}) \,
\mathcal{K}(\mathbf{q_1},\mathbf{q_2},z) \,
\tilde{\delta}^{(1)}_{0}(\mathbf{q_1}) \,
\tilde{\delta}^{(1)}_{0}(\mathbf{q_2}) \;,
\end{equation}
where 
$D_+(z)$ is the linear grow factor of density fluctuations 
and $\delta^{(1)}_0$ is the matter overdensity
linearly extrapolated till the present epoch (i.e. at $z=0$ when $D_+=1$). 
For the density field originated via gravitational instability of Gaussian initial conditions,
the kernel $\mathcal{K}$ assumes the form (e.g. Fry 1984)
\be
\label{GRAVITATIONAL}
\mathcal{K}_{\rm L}(\mathbf{k_1},\mathbf{k_2}) = 
\frac{5}{7}+\frac{2}{7}\cos^2\theta_{12} + 
\frac{1}{2} \left(\frac{k_1}{k_2}+\frac{k_2}{k_1} 
\right)\cos\theta_{12}
\ee
with $\cos\theta_{12}=\hat{\mathbf k}_1\cdot\hat{\mathbf k}_2$ and $\hat{\mathbf k}={\mathbf k}/k$.
This holds at high-redshift (where the mass density parameter approaches
unity). 

 \subsection{Primordial non-Gaussianity}
\label{NG}

For a large class of models for the generation of the initial seeds
for structure formation, including standard single-field and multi-field 
inflation, the curvaton and the inhomogeneous reheating scenarios, 
the level of primordial non-Gaussianity can be modeled through a 
quadratic term in Bardeen's gauge-invariant potential $\Phi$, 
\footnote{On scales much smaller than the Hubble radius, Bardeen's 
gauge-invariant potential would reduce to minus the usual peculiar 
gravitational potential.} namely 
\begin{equation}
\label{FNL}
\Phi = \Phi_{\rm L} + f_{\rm{NL}} \left(\Phi_{\rm L}^2 - 
\langle\Phi_{\rm L}^2\rangle \right) \;,
\end{equation}
where $\Phi_{\rm L}$ is a Gaussian random field and the specific value of
the dimensionless non-linearity parameter $f_{\rm{NL}}$ depends on the assumed
scenario (e.g. Bartolo et al. 2004). \nocite{BMKR04} 
The kernel 
\be
\label{KERNEL}
\mathcal{K}(\mathbf{q_1},\mathbf{q_2},z) = 
\mathcal{K}_{\rm L}(\mathbf{q_1},\mathbf{q_2})+
\mathcal{K}_{\rm NL}(\mathbf{q_1},\mathbf{q_2},z)
\ee
with
\be
\label{PRIMORDIAL}
\mathcal{K}_{\rm NL}(\mathbf{q_1},\mathbf{q_2},z)=
6\, f_{\rm{NL}}\,  E(z) \, \frac{T(|{\bf q}_1 + {\bf q}_2|)}{T(q_1)\,T(q_2)}\,
\left( \frac{1}{q^2_1}+\frac{1}{q^2_2} + 
\frac{2}{q_1 q_2}\cos\theta_{12}  \right)
\ee
accounts for both the non-linear effects of gravitational instability ($\mathcal{K}_{\rm L}$) and the 
evolved non-Gaussian contribution ($\mathcal{K}_{\rm NL}$). 
Here $E(z)=H_0^2\,\Omega_{0m}/(4\,c^2\,D_+(z))$,
and $T(k)$ is the linear transfer function for CDM.

It is worth stressing that equation (\ref{FNL}), even though commonly used, 
is not generally valid: 
detailed second-order calculations of the evolution of perturbations from the inflationary period to the 
present time show that the quadratic, 
non-Gaussian contribution to the gravitational potential should be 
represented as a convolution with a kernel $f_{\rm NL}({\bf x},{\bf y})$ 
rather than a product (in full analogy with the quadratic term 
in equation (\ref{GUNOGDUERS})). Indeed, for scales which entered 
the Hubble radius during matter dominance, the 
calculation of $f_{\rm NL}$ can be performed analytically in the
so-called Poisson gauge (Bartolo, Matarrese \& Riotto 2005) and leads to 
the Fourier space expression: 
\begin{eqnarray}
\tilde{f}_{\rm NL}({\bf k}_1,{\bf k}_2;z)&=&
\frac{5}{3}\, \left(a_{\rm nl}-1\right)
+\frac{77({\bf k}_1 \cdot {\bf k}_2) - 102(k_1^2+k_2^2)}{42 k^2}+\nonumber\\
&-& \frac{(k_1^2+k_2^2)({\bf k}_1 \cdot {\bf k}_2)}{k_1^2 k_2^2 k^2}\,
\left(\frac{4}{7}({\bf k}_1 \cdot {\bf k}_2)
+k_1^2+k_2^2\right)\nonumber \\
&+&\frac{6}{7}\frac{k_1^2k_2^2}{k^4} - \frac{6}{7}
\frac{({\bf k}_1 \cdot {\bf k}_2)^2}{k^4} + {\cal O}(E(z)^2/k^4) \;, 
\label{fnlcomplete}
\end{eqnarray}
where the parameter $a_{\rm nl}-1$ quantifies the amount of 
non-Gaussianity produced during (or immediately after) inflation.
For the simplest case of single-field slow-roll inflation, 
$|a_{\rm nl}-1| \ll 1$, being of the order of the slow-roll parameters 
(Gangui et al. 1994; Acquaviva et al. 2003; 
Maldacena 2003), while alternative models, like the curvaton or the 
inhomogeneous reheating scenarios, may easily accomodate much larger values 
of $a_{\rm nl}-1$ (see Bartolo et al. 2004 and references therein), 
such that all momentum-dependent contributions to $\tilde{f}_{\rm NL}$
can be neglected and the non-linearity parameter can be effectively 
approximated with a constant, $\tilde{f}_{\rm NL} \approx \frac{5}{3} (a_{\rm nl}-1)$, 
as assumed in the present analysis (see equation (\ref{FNL})).   

From the expressions above, one can easily obtain the (tree-level)
bispectrum of density perturbations 
evaluated at three different redshifts as 
\be 
\label{213DBISPVSPOWER}
B(\mathbf{k_1},\mathbf{k_2},\mathbf{k_3},z_1,z_2,z_3) = 
2~\mathcal{K}(\mathbf{k_1},\mathbf{k_2},z_1) D_+(z_1)^2 D_+(z_2) D_+(z_3)
P_0(k_1) P_0(k_2) + \mathrm{cycl.}\;,
\ee
where $P_0(k)$ is the linear power-spectrum of density fluctuation extrapolated
till the present epoch.  
The above expressions tell us that the optimal strategy to constrain 
non-Gaussianity in the large-scale structure of the Universe is to 
use large-scale and/or high-redshifts datasets (e.g. Verde, Heavens
\& Matarrese 2000b; Verde et al. 2001;
Scoccimarro, Sefusatti \& Zaldarriaga 2004). 
From this point of view, the 21-cm fluctuations potentially represent a 
very powerful tool to detect non-Gaussianity which could complement 
the limits 
obtained from the study of CMB anisotropies (e.g. Spergel et al. 2006). 

\section{The angular bispectrum of 21-cm fluctuations}
\label{calc}
\subsection{Definition}
\label{BISP}

The observed brightness temperature is a convolution of $\TBRIGHT$ 
with a frequency-dependent instrumental response
$W_{\nu}(r)$ (from now on the window function):
\begin{equation}
\TBRIGHT^{\rm obs}(\hat{\bf n})= \int d r ~ W_{\nu}(r)~  T_{\rm b}(r\hat{\bf n}).
\end{equation}
The window function peaks at the central frequency of the detector, $\nu$, 
and is different from zero within a characteristic frequency range 
corresponding to the bandwidth of the instrument. 
For analytic convenience we prefer here to express $W_{\nu}$ 
in terms of the comoving distance $r$ evaluated at $z=\nu_{21}/\nu-1$.

The fluctuations of the observed 
\footnote{In order to simplify the notation, from now on we drop the 
superscript ``obs'' to indicate the observed brightness temperature.}
brightness temperature can be expanded in spherical harmonics:
\begin{equation}
\Delta \TBRIGHT(\hat{\bf n})= \sum_{\ell=0}^{\infty} \sum_{m=-\ell}^{\ell} a_{\ell m} Y_{\ell m}(\hat{\bf n}),
\ \ \ \ \ 
a_{\ell m}= \int d^2\hat{\bf n} \,\Delta \TBRIGHT(\hat{\bf n})\,  Y^{\ast}_{\ell m}(\hat{\bf n})\;.
\end{equation}
The angular bispectrum is then given by the ensemble average
\begin{equation}
\label{ANGBISP}
B_{\ell_1 \ell_2 \ell_3}^{m_1 m_2 m_3} =~ \langle a_{\ell_1 m_1} a_{\ell_2 m_2} a_{\ell_3 m_3}\rangle\;,
\end{equation}
which for an isotropic random field can be re-written in terms of the
angle-averaged bispectrum $B_{\ell_1 \ell_2 \ell_3}$ as
\begin{equation}
B_{\ell_1 \ell_2 \ell_3}^{m_1 m_2 m_3}=  \left( \begin{array}{ccc}
\ell_1&\ell_2&\ell_3\\
m_1&m_2&m_3\\
\end{array} \right) B_{\ell_1 \ell_2 \ell_3}\;.
\end{equation}
Here the matrix denotes the Wigner-$3j$ symbol so that $B_{\ell_1 \ell_2 \ell_3}^{m_1 m_2 m_3}$
does not vanish only when:
\begin{enumerate}
\item $\left| \ell_j -\ell_k \right| \leq \ell_{\rm{i}} \leq \ell_j + \ell_k$ for all permutations of indices $i,j,k\;$;
\item $\ell_1+\ell_2+\ell_3 =$ even$\;$;
\item $m_1+m_2+m_3 = 0\;$.
\end{enumerate}

\subsection{Derivation}

The angular bispectrum of the 21-cm background is then
\footnote{We just consider the case of a single window function.
The generalization of our results to multi-frequency analyses is straightforward.}
\begin{eqnarray}
\label{bispaaa}
\langle a_{\ell_1 m_1}~ a_{\ell_2 m_2}~ a_{\ell_3 m_3} \rangle~&=&
\int d\hat{\bf n}_1~ d\hat{\bf n}_2~ d\hat{\bf n}_3 \int  dr_1~  dr_2 ~dr_3~ W_{\nu}(r_1) W_{\nu}(r_2) W_{\nu}(r_3) ~  \\
\nonumber
& &  \langle \Delta \TBRIGHT(\mathbf{r_1})~ \Delta \TBRIGHT(\mathbf{r_2}) ~\Delta \TBRIGHT(\mathbf{r_3}) \rangle ~ Y^{\ast}_{\ell_1 m_1}(\hat{\bf n}_1)~Y^{\ast}_{\ell_2 m_2}(\hat{\bf n}_2)~Y^{\ast}_{\ell_3 m_3}(\hat{\bf n}_3)\;,
\end{eqnarray}
where the three-point correlation function is
%
\begin{eqnarray}
\label{3DBISP}
\langle \Delta \TBRIGHT(\mathbf{r_1})~\Delta \TBRIGHT(\mathbf{r_2})~\Delta \TBRIGHT(\mathbf{r_3})\rangle ~&=&
\int \frac{d^3\mathbf{k}_1~ d^3 \mathbf{k}_2~ d^3 \mathbf{k}_3}{(2\pi)^6}~ B_{\rm{21}}(\mathbf{k}_1,\mathbf{k}_2,\mathbf{k}_3,z_1,z_2,z_3)\,\\ 
\nonumber
& &\delta_{\rm{D}}(\mathbf{k}_1+\mathbf{k}_2+\mathbf{k}_3) 
e^{i\,( \mathbf{k_1}  \mathbf{r_1}+\mathbf{k_2}  \mathbf{r_2}+\mathbf{k_3}  \mathbf{r_3})}
\end{eqnarray}
%
with $B_{21}(\mathbf{k_1},\mathbf{k_2},\mathbf{k_3},z_1,z_2,z_3)$ the three-dimensional bispectrum of $\Delta\TBRIGHT$.
\subsubsection{Bispectrum from gravitational instability}
We first account for the non-Gaussianity generated by gravitational instability from Gaussian initial conditions.
In this case, 
the integral over ${\bf k_3}$ in equation (\ref{3DBISP})
can be easily computed exploiting the Dirac delta function.
We then expand the exponential term in the same equation 
into spherical harmonics and perform all the angular integrations.
On scales where the baryon distribution traces the density fluctuations of the dark matter, 
\begin{equation}
B_{21}(\mathbf{k_1},\mathbf{k_2},\mathbf{k_3},z_1,z_2,z_3) = 
2~\mathcal{K}_{21}(\mathbf{k_1},\mathbf{k_2},z_1) D_+(z_1)^2 D_+(z_2) D_+(z_3) f_1(z_1) f_1(z_2) f_1(z_3)
P_0(k_1) P_0(k_2) + \mathrm{cycl.}\;.
\end{equation}
where the kernel $\mathcal{K}_{\rm{21}}$ can be written as a power series of $\cos\theta_{12}$,
\begin{equation}
\label{cosineseries}
2\,\mathcal{K}_{\rm{21}}(\mathbf{k_1},\mathbf{k_2},z) = A_{\rm{0}} + A_1~ \cos\theta_{12} + A_2~\cos^2\theta_{12}
\end{equation}
with
\begin{equation}
\label{acoeff}
A_{\rm{0}} = \frac{10}{7} + \frac{f_3}{f_1}~g_{\rm 2a} 
+\frac{f_2}{f_1}\;, \ \ \ \ \ \ \
A_1 = \left(1+\frac{f_3}{f_1}~g_{\rm 2c} \right)\, 
\left( \frac{k_1}{k_2}+\frac{k_2}{k_1} \right)\;, \ \ \ \ \ \ \
A_2 = \frac{4}{7}+\frac{f_3}{f_1}~g_{\rm 2b}\;.
\end{equation}
The factors $g_{\rm 2a},g_{\rm 2b},g_{\rm 2c}$ 
are functions of time that parameterize the evolution of the kernel $\GDUE$
and thus take into account gas temperature 
perturbations up to second order in $\delta$ as indicated in Appendix \ref{PGT}.
After long but straightforward calculations (see Verde et al. 2000b 
for a similar case), \nocite{VERHEAMAT}
the expression for the angular bispectrum of the 21-cm background can be written as a sum of cyclic terms
\begin{equation}
B_{\ell_1 \ell_2 \ell_3} = B_{\ell_1 \ell_2} + B_{\ell_1 \ell_3} + B_{\ell_2 \ell_3}
\end{equation}
with
\begin{eqnarray}
\label{21BISP}
B_{\ell_1 \ell_2} &=& \frac{16}{\pi} ~ i^{(\ell_1+\ell_2)}\, \sqrt{ \frac{(2\ell_1+1)(2\ell_2+1)(2\ell_3+1)}{{(4\pi)}^3} } \int dk_1 \int dk_2~ k^2_1 \,  k^2_2\,  \\
\nonumber
& & \alpha_{\ell_1}(k_1)\,   \alpha_{\ell_2}(k_2)\, zP_0(k_1)\,  P_0(k_2)
\sum_{\ell,\ell_6,\ell_7} i^{(\ell_6+\ell_7)}(-1)^{\ell}     (2\ell_6+1)(2\ell_7+1) \\
\nonumber
& &  \left( \begin{array}{ccc}
\ell_1&\ell_6&\ell\\
0&0&0\\
\end{array} \right) 
\left( \begin{array}{ccc}
\ell_2&\ell_7&\ell\\
0&0&0\\
\end{array} \right)
 \left( \begin{array}{ccc}
\ell_3&\ell_6&\ell_7\\
0&0&0\\
\end{array} \right)
 \left \{ \begin{array}{ccc}
\ell_1&\ell_2&\ell_3\\
\ell_7&\ell_6&\ell\\
\end{array} \right \}
\\
\nonumber
& &  \int dr   D_+(r)^2\, f_1(r)\, W_{\nu}(r)\,  j_{\ell_6}(k_1 r)\,   j_{\ell_7}(k_2 r)\, B_\ell(k_1,k_2,r)\;, 
\end{eqnarray}
where $\{\dots\}$ indicates a Wigner-$6j$ symbol. 
Here $\alpha_{\ell}(k)=\int dr\,  j_{\ell}(k r)\,  D_+(r)\,f_1(r)\,W_{\nu}(r)$, 
with $j_{\ell}(x)$ the spherical Bessel function of the first kind.
Note that only terms with $\ell_6=\ell_1-\ell, .., \ell_1+\ell$ and $\ell_7=\ell_2-\ell, .., \ell_2+\ell$ contribute to the sum.
The functions $ B_\ell(k_1,k_2,r)$ are generated by the terms in equation (\ref{cosineseries}). 
Since this power series does not contain exponents larger than 2, only terms with $\ell=0,1,2$ matter. 
We thus obtain:
\begin{eqnarray}
B_{0} &=& A_0+\frac{1}{3}A_2 = \frac{34}{21}+\frac{f_3}{f_1}~\left(g_{\rm 2a}+\frac{1}{3}~g_{\rm 2b} 
\right)+\frac{f_2}{f_1}\;, \\
B_1 &=& A_1 = (1+\frac{f_3}{f_1}~g_{\rm 2c}) \left(\frac{k_1}{k_2}+\frac{k_2}{k_1} \right)\;, \\
B_2 &=& \frac{2}{3}A_2 =\frac{8}{21}+\frac{2}{3}\frac{f_3}{f_1}~g_{\rm 2b}\;.
\end{eqnarray}
For narrow window functions, the expression in equation (\ref{21BISP}) can be further simplified. 
In particular, the linear growth factors, the coupling constants $f_i$ and the time functions $g_{\rm 2i}$ 
introduced in Appendix \ref{PGT} can be evaluated at the epoch corresponding to the central frequency 
of the detector.

\subsubsection{Bispectrum from primordial non-Gaussianity}
We follow a different procedure to integrate equations (\ref{bispaaa}) and (\ref{3DBISP}) with the 
kernel $\mathcal{K}_{\rm NL}(\mathbf{q_1},\mathbf{q_2},z)$ given in equation (\ref{PRIMORDIAL}).
In this case, it is convenient to expand the Dirac delta function in spherical harmonics and perform the angular
integrations (e.g. Komatsu \& Spergel 2001).
The result is
\be
\label{bispnl}
B_{\ell_1 \ell_2 \ell_3}=
\frac{16}{\pi^3}\,f_{\rm NL}\,\sqrt{\frac{(2\ell_1+1)(2\ell_2+1)(2\ell_3+1)}{4\pi}} \,\left( \begin{array}{ccc}
\ell_1&\ell_2&\ell_3\\ 0&0&0\\
\end{array} \right) 
\int dr\,r^2\,\left[\beta_{\ell_1}(r)\,\beta_{\ell_2}(r)\,\Gamma_{\ell_3}(r) + {\rm cycl.}\right]
\ee
where
\be
\beta_\ell(r)= \int dk\,\frac{P_0(k)}{T(k)}\,\alpha_\ell(k)\,j_\ell(kr)
\ \ \ \ \ \ \ {\rm and} \ \ \ \ \ \ \
\Gamma_\ell(r)= 6\,\int dk\,k^4\,T(k)\,\epsilon_\ell(k)\,j_\ell(kr)\;,
\ee
with $\epsilon_\ell(k)=\int dr'\, D_+(r')^2\,[E(r')\,f_1(r')+g_{\rm 2d}(r')f_3(r')/6]\,
W_{\nu}(r')\,j_\ell(kr')$.
Note that, at $z\simeq 50$, $6 E f_1 / g_{\rm 2d} f_3\simeq 20$
and the contribution of the non-local kernel $\GDUE$ to $\epsilon_\ell$ is rather small
(see also Figure \ref{KERNELSFIG}).

\subsection{Results}

\begin{figure}\epsscale{.9}
\plottwo{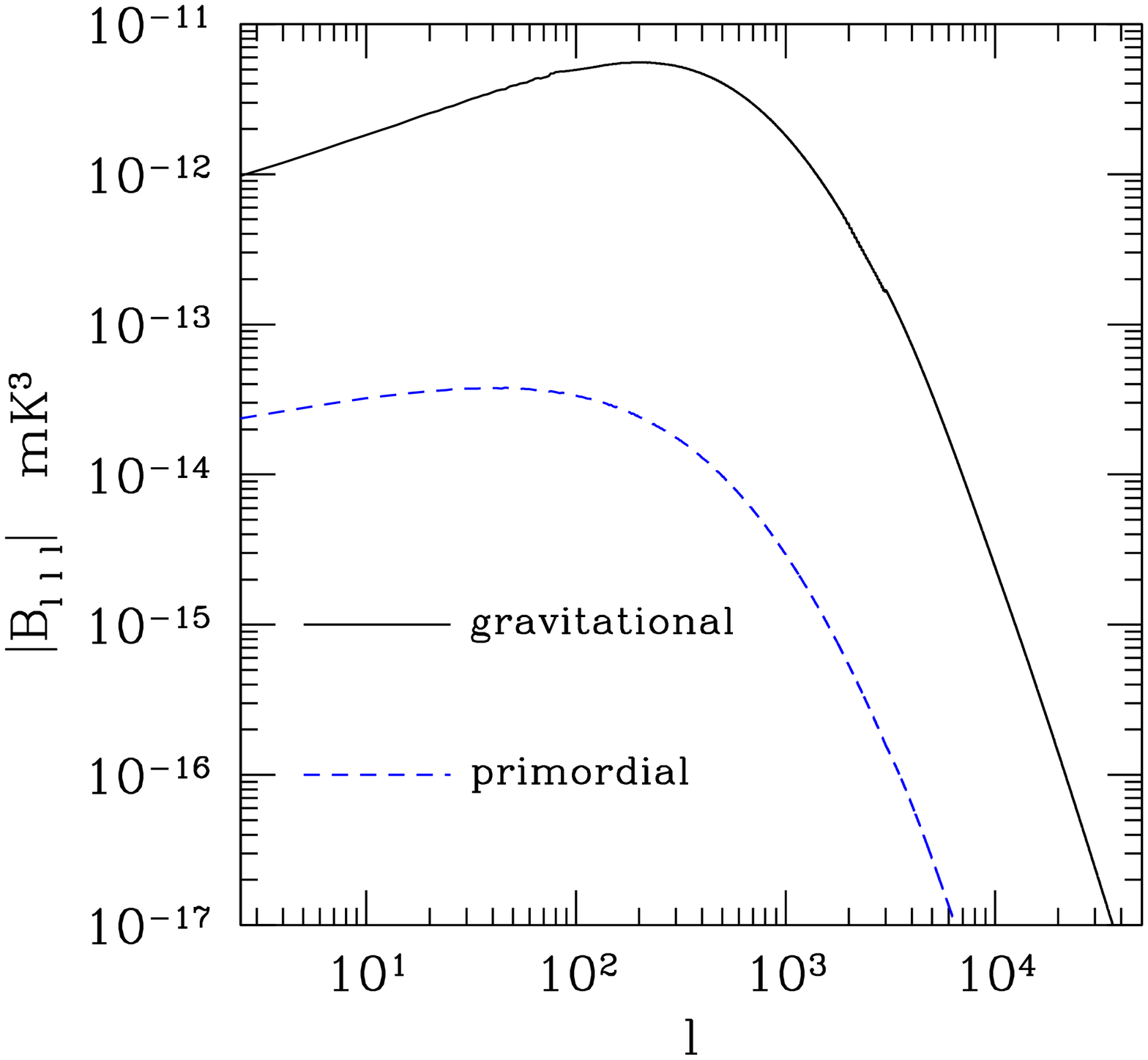}{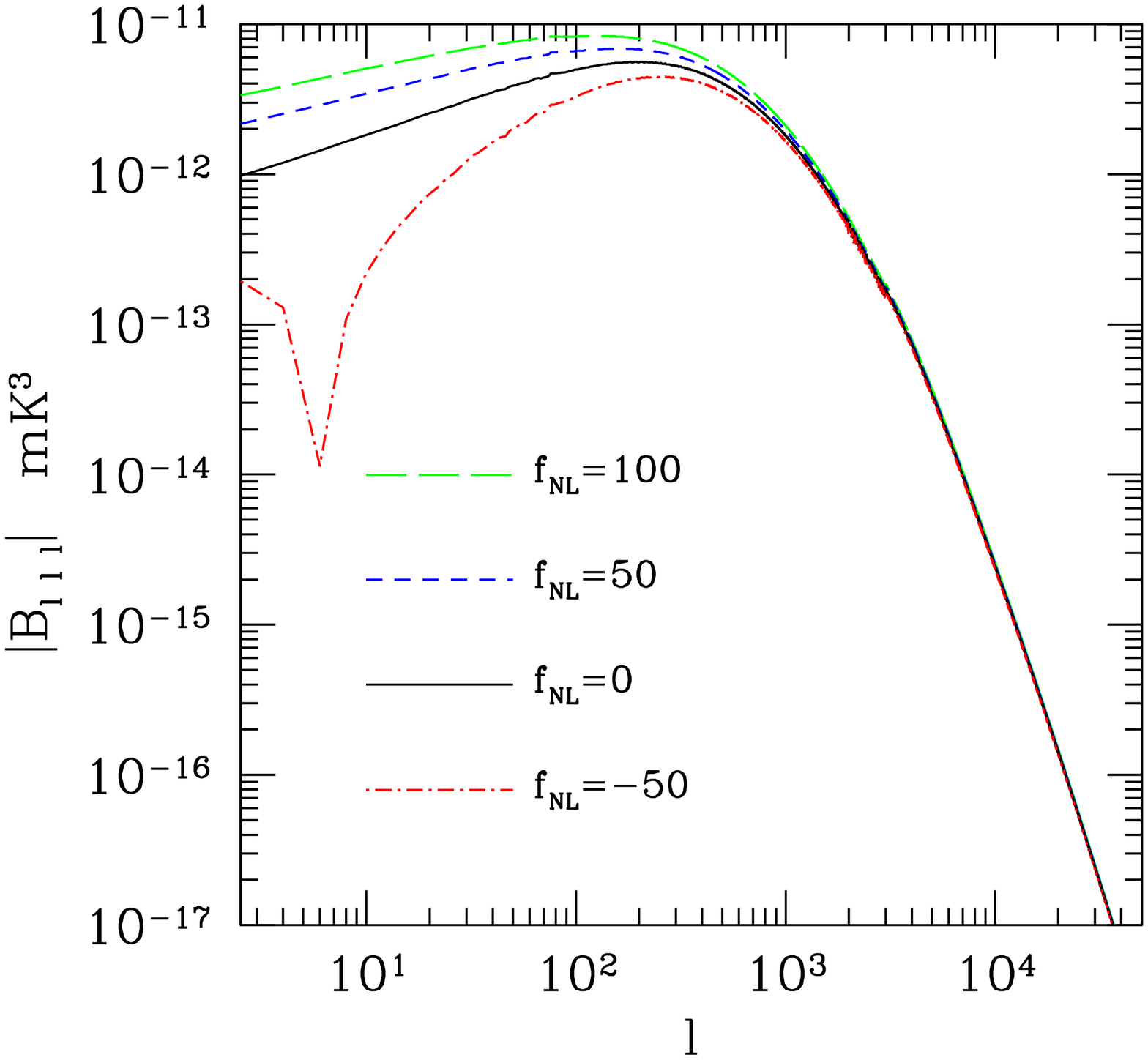}
\plottwo{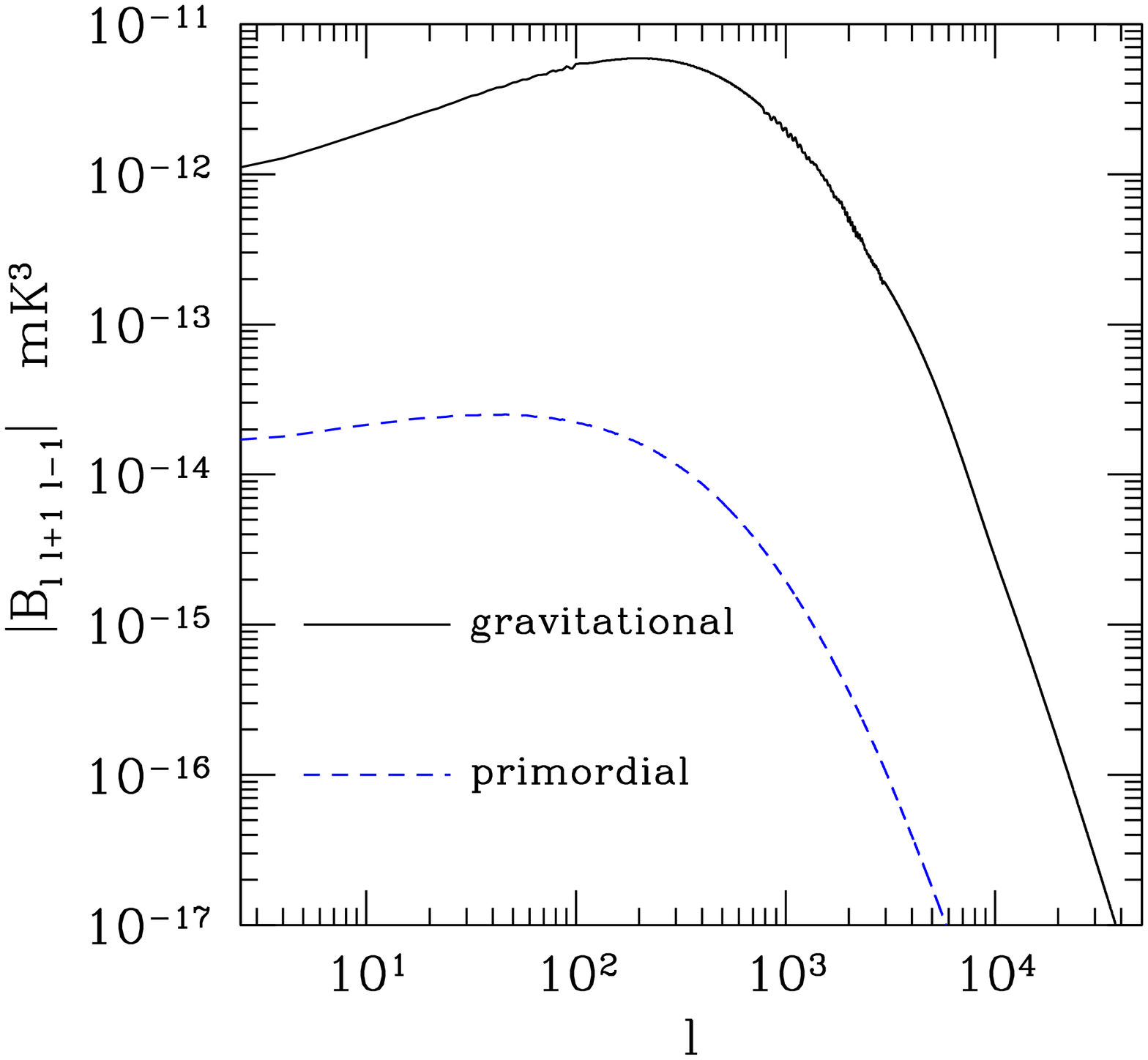}{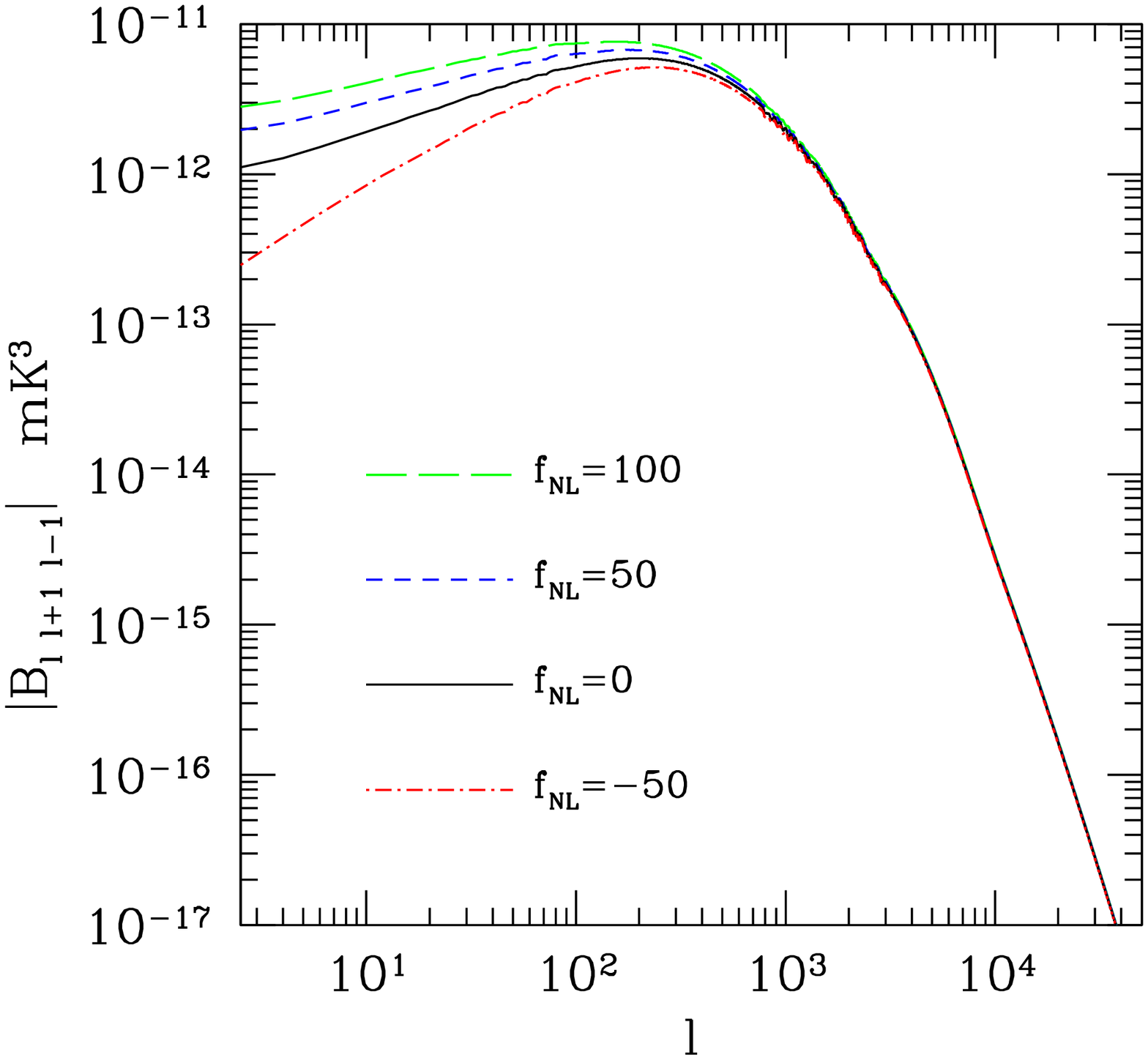}
\plottwo{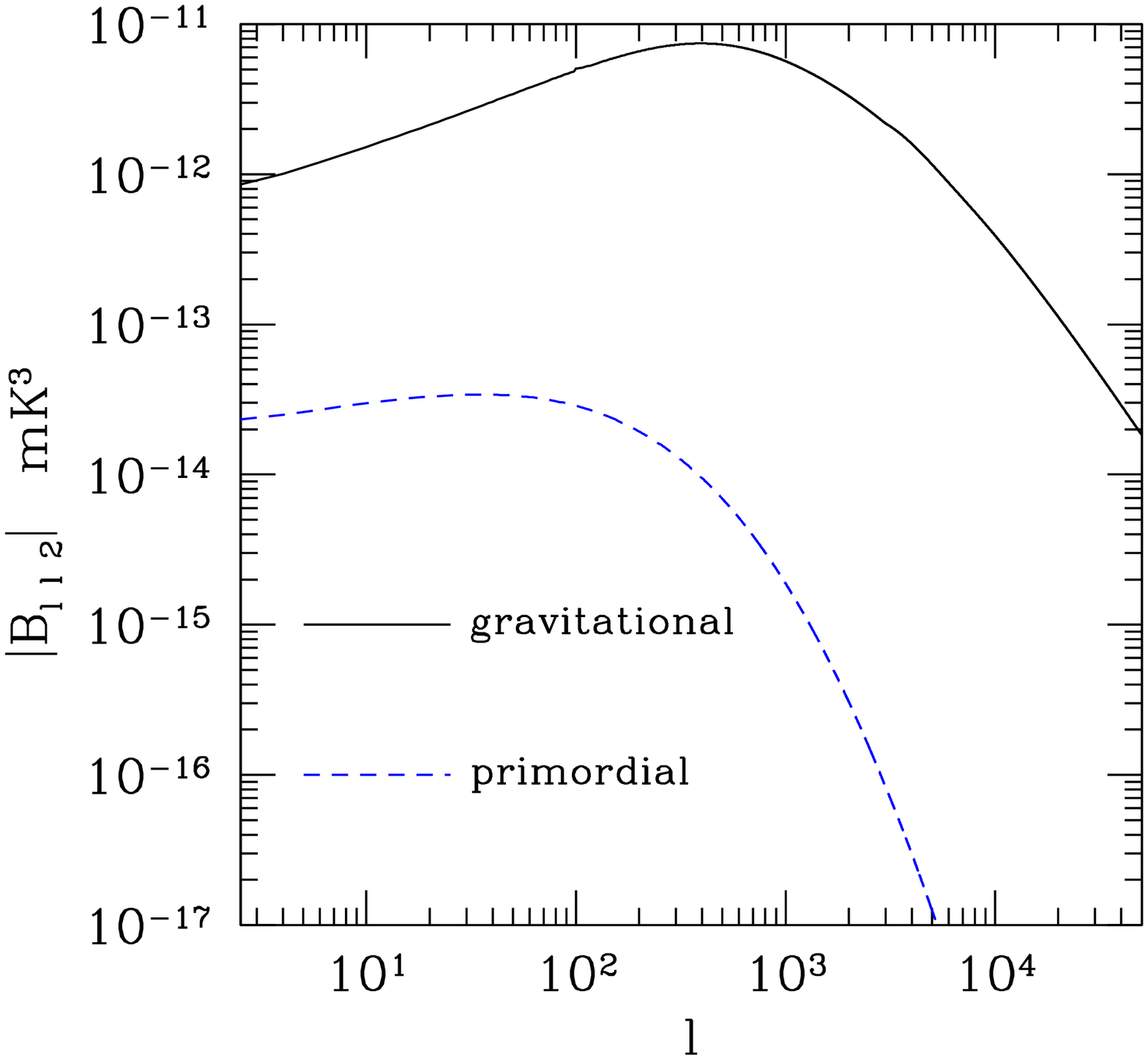}{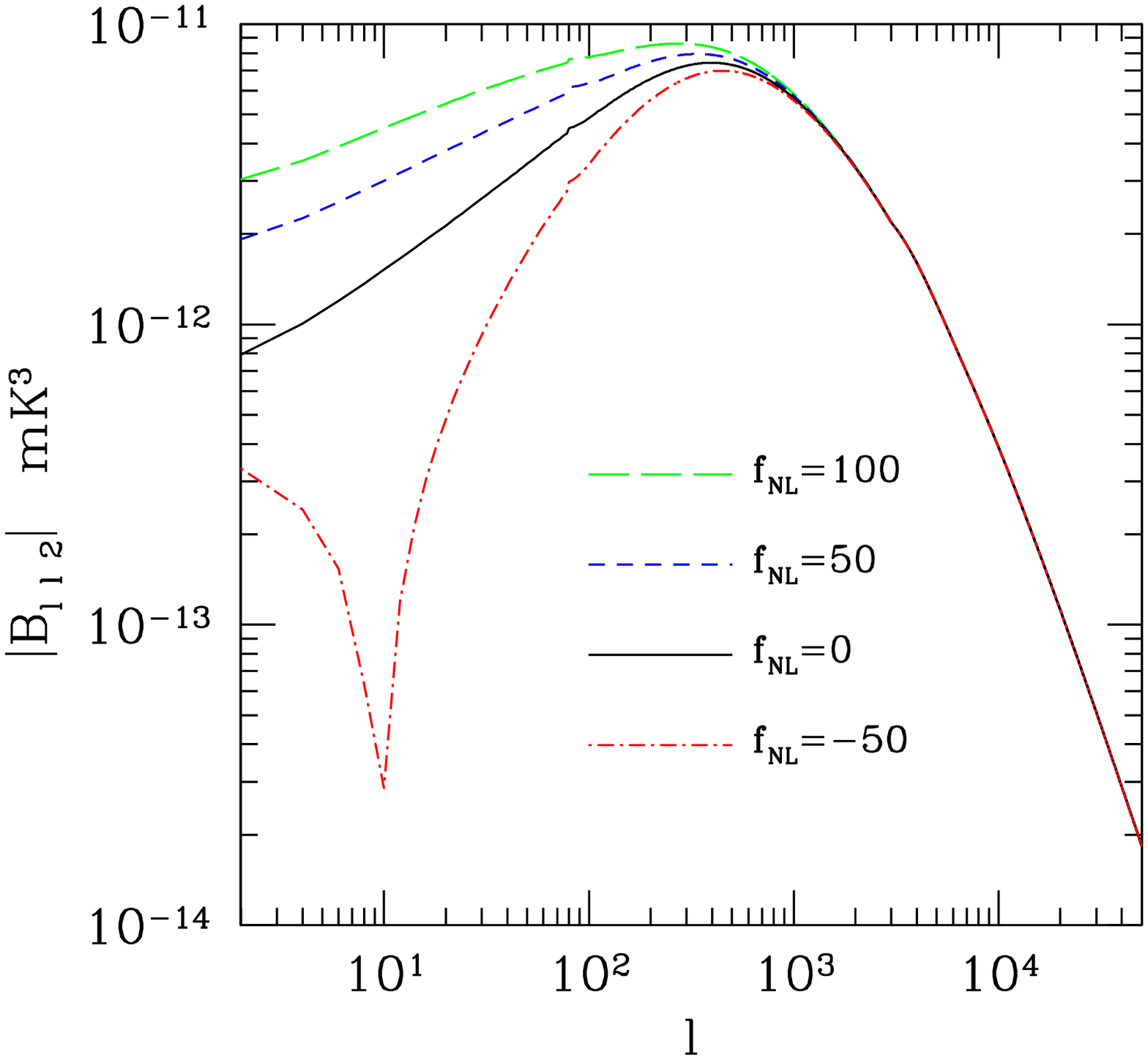}
\epsscale{1}
\caption{\label{BISPFIG}\small{The bispectrum  
of 21-cm anisotropies $B_{\ell_1,\ell_2,\ell_3}$ measured by an ideal
experiment with 0.1 MHz bandwidth centered around $z=50$. 
From top to bottom, we consider
equilateral ($ \ell_1= \ell_2= \ell_3$), quasi-equilateral
($\ell_2=\ell_1+1,\ell_2=\ell_1-1$) and ``squeezed'' 
($\ell_2=\ell_1, \ell_3=2$) configurations, respectively.
In the left panels,
the contributions of non-linear gravity (solid) and of primordial
non-Gaussianity (with $f_{\rm NL}=1$; dashed) are compared. 
The right panels show
the total bispectrum for different values
of the non-linearity parameter. Note that in the equilateral and quasi-equilateral
configurations the bispectrum vanishes when $\ell$ is odd (not plotted), is positive
for $\ell=2j$ with $j$ an odd integer and negative for $\ell=2j$ with $j$ an even integer. }}
\end{figure}

We perform a numerical integration of 
equations (\ref{21BISP}) and (\ref{bispnl}).
The resulting angular bispectra are shown in
Figure \ref{BISPFIG} for different sets of the parameters $\ell_1,
\ell_2$ and $\ell_3$ (bispectrum configurations).
From top to bottom, we present equilateral ($B_{\ell\, \ell\, \ell}$),
quasi-equilateral ($B_{\ell\,\ell+1\, \ell-1}$) and ``squeezed'' 
($B_{\ell\, \ell\, 2}$) configurations. 
In all cases, we consider an ideal instrument  with unlimited angular 
resolution and characterized by a Gaussian window function 
with a frequency ($1\sigma$) bandwidth of 0.1 MHz centred around $z=50$.
The left panels compare the different contributions due to non-linear gravity
-- equation (\ref{21BISP}) -- and to primordial non-Gaussianity 
-- equation (\ref{bispnl}) -- with $f_{\rm NL}=1$.
The total non-Gaussian signal in the various configurations and 
for different values of $f_{\rm NL}$ is shown in the right panels
of Figure \ref{BISPFIG}.
Note that the signal due to gravitational instability dominates at all angular
scales. For $\ell<200$, the ratio between the two contributions is $\sim 50$.
This difference becomes much more severe at smaller angular scales. 
This happens because any primordial non-Gaussianity in the gravitational 
potential is going to be shifted to larger scales when observed in the 
overdensity field as a direct consequence of the Poisson equation.

\section{Discussion and conclusions}
\label{concl}

\begin{figure}
\plottwo{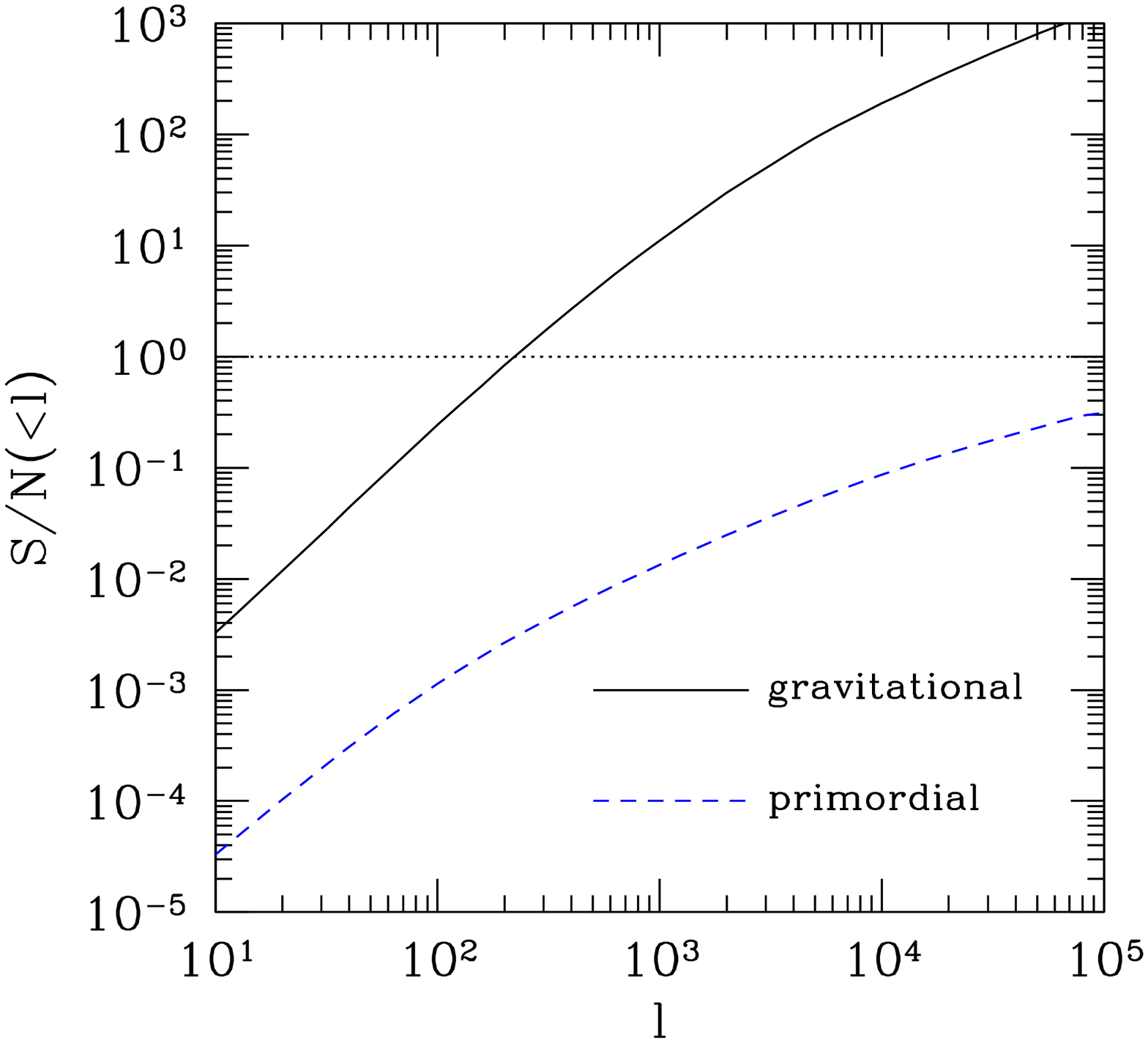}{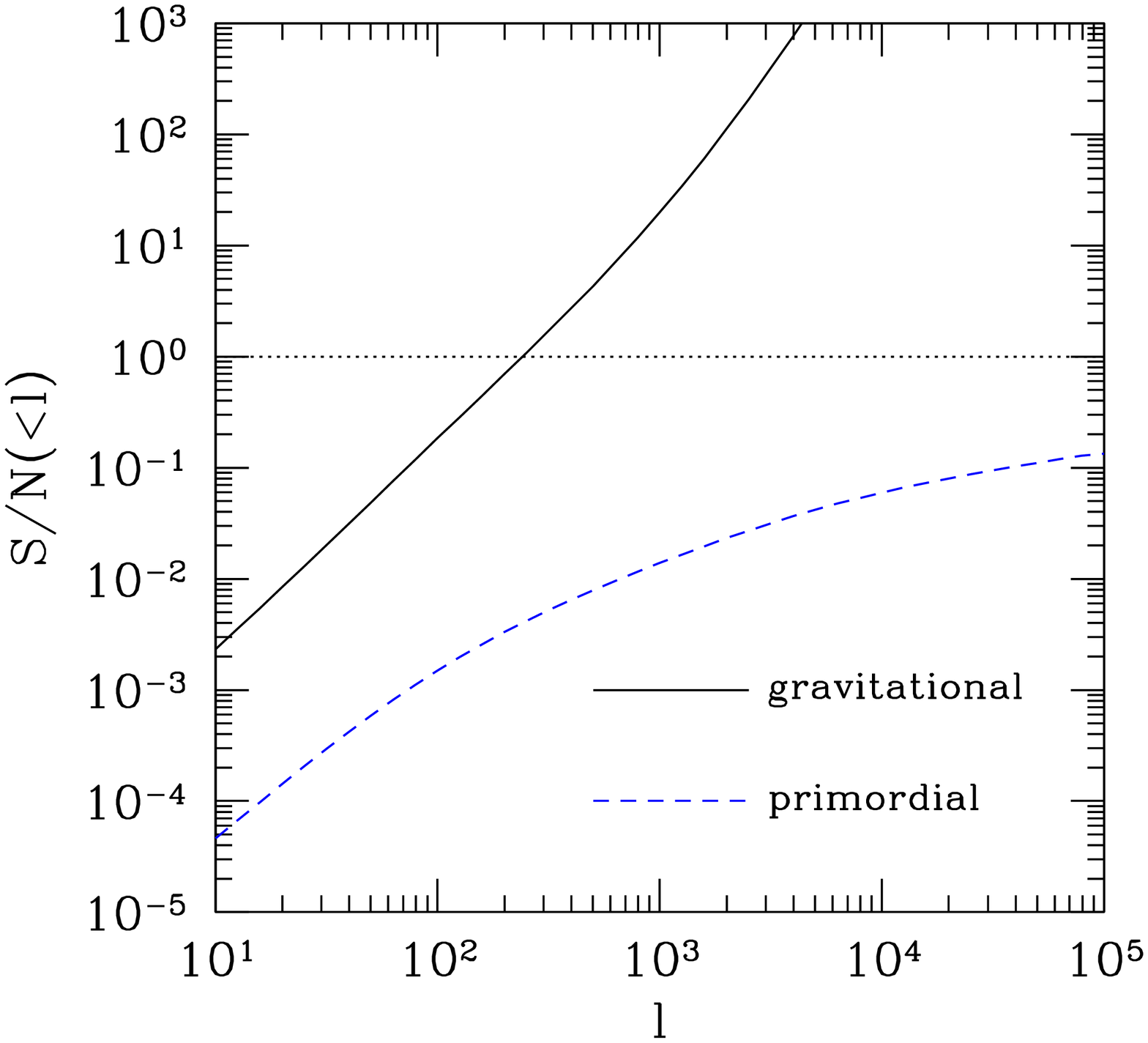}
\caption{\label{SNFIG}
Cumulative signal-to-noise ratio for the measurement of the bispectrum
of 21-cm anisotropies using modes up to a maximum value of $\ell$.
Solid lines refer to non-Gaussianity generated by gravity while dashed lines
indicate the primordial signal with $f_{\rm NL}=1$. 
These quantities have been computed using the signal-to-noise ratio per mode of
the quasi-equilateral configuration (left) and of the squeezed configuration
(right).}
\end{figure}

Using second order perturbation theory, we have computed analytically and
evaluated numerically the bispectrum of redshifted 21-cm fluctuations down to 
a few arcsecond angular scales ($\ell\simeq 10^5$).
Is the expected signal detectable by future radio experiments? And can the data be used to test the gravitational instability
scenario and quantify primordial non-Gaussianity?
Suppose we fit the observed bispectrum $B_{\ell_1\ell_2\ell_3}^{\rm obs}$ 
with a theoretical model by minimizing the objective function
\be
\chi^2=\sum_{\ell_1\leq\ell_2\leq\ell_3}
\frac{\left(B_{\ell_1\ell_2\ell_3}^{\rm obs}-\sum_i P_i\, B^{(i)}_{\ell_1\ell_2\ell_3}\right)^2}{\sigma^2_{\ell_1\ell_2\ell_3}}
\ee
where $i$ distinguishes the different contributions of non-linear gravity and primordial non-Gaussianity and $P_i$ is a normalization constant 
(i.e., for a given set of cosmological parameters, $P_1\equiv (\sigma_8/0.9)^4$ 
and $P_2\equiv f_{\rm NL}\,(\sigma_8/0.9)^4$).
If the amount of non-Gaussianity is small, the cosmic variance of the bispectrum is given by the disconnected
part of the six-point function of $a_{\ell m}$ (Luo 1994). 
The variance of $B_{\ell_1\ell_2\ell_3}$ is then calculated as (e.g. 
Spergel \& Goldberg 1999)
\be
\sigma^2_{\ell_1\ell_2\ell_3}=\langle B^2_{\ell_1\ell_2\ell_3} \rangle - \langle B_{\ell_1\ell_2\ell_3} \rangle^2 \simeq {\mathcal C}_{\ell_1}\,{\mathcal C}_{\ell_2}\,
{\mathcal C}_{\ell_3}\,\Delta_{\ell_1\ell_2\ell_3}\;,
\ee
where $\Delta_{\ell_1\ell_2\ell_3}$ is equal to 1, 2 or 6 when all the $\ell$-indices are different, two of them are the same
or all of them are identical, respectively. Here ${\mathcal C}_{\ell}$ denotes the total angular power spectrum of the 21-cm background including
the contribution of the detector noise.
The corresponding Fisher matrix $F_{ij}$ is
\be
F_{ij}=\sum_{\ell_1\leq\ell_2\leq\ell_3} \frac{B_{\ell_1\ell_2\ell_3}^{(i)} B_{\ell_1\ell_2\ell_3}^{(j)}}
{\sigma^2_{\ell_1\ell_2\ell_3}}\;,
\ee
where $i=1$ refers to parameter $P_1$ and $i=2$ refers to parameter $P_2$. The signal-to-noise ratio $(S/N)_i$ for the $i$-component is given by 
\be
\left(\frac{S}{N} \right)_i=\frac{1}{\sqrt{F_{ii}^{-1}}}\;.
\ee
An order of magnitude estimation of this quantity as a function of the angular resolution $\ell$ 
can be obtained neglecting the covariance of the different components, thus writing $(S/N)_i\simeq \sqrt{F_{ii}}$.
Since the calculation of $B_{\ell_1\ell_2\ell_3}$ is expensive at high $\ell$, we use the configurations in Figure
\ref{BISPFIG} to estimate the total signal-to-noise ratio. Our results are shown in Figure \ref{SNFIG}: an experiment with arcmin-scale resolution
at $z=50$ would clearly detect the bispectrum induced by non-linear gravity with $(S/N)\sim 100$. 
On the other hand, for the signal from primordial non-Gaussianity, 
$(S/N)\sim 0.1\, f_{\rm NL}$.
Increasing the angular resolution by a factor of 10 would give
$(S/N)\sim 0.3\, f_{\rm NL}$.
We consider here an ideal, full-sky experiment where the measurements are only limited by cosmic variance 
(i.e. with no detector noise) and where foreground signals can be perfectly subtracted. 
If only a fraction $f_{\rm sky}$ is surveyed, our signal-to-noise estimates should be depressed by a factor $f^{1/2}_{\rm sky}$.

A tomography of the neutral hydrogen distribution within 
the redshift range $30<z<100$ in slices of bandwidth 
$\sim 0.1$ MHz ($1\sigma$)
would provide nearly 150 21-cm maps. Since the correlations 
between different redshift slices will be negligible for high multiples ($\ell>r/\Delta r\simeq 1500$), 
this would increase the signal-to-noise ratios by a factor of $\sim 10$.

Our results show that studies of the 
21-cm bispectrum have the potential to test the gravitational instability scenario and the origin of primordial fluctuations.
Measurements of three-point statistics will also provide information on the cosmological parameters
through the shape dependence of $B_{\ell_1\ell_2\ell_3}$ and the redshift evolution of the growth factors. 

Low-frequency radio observations with high-angular resolution (a few arcsec,
corresponding to $\ell\sim 10^5$)
can detect primordial non-Gaussianity with $f_{\rm NL}\sim 1$.
One should however take into account that such high values of $\ell$
correspond to perturbation modes which have crossed the Hubble radius
during radiation dominance; for such modes and for values of
$f_{\rm NL} \sim 1$, a full second-order evaluation of the
matter-transfer function would be needed to account for e.g. the
M\'esz\'aros effect beyond the linear approximation, following Bartolo,
Matarrese \& Riotto (2007).
On the scales of interest, weak lensing by the large-scale structure
of the Universe typically modifies the angular power spectrum of 21-cm 
fluctuations
by only $\sim 1\%$ (Mandel \& Zaldarriaga 2006). However, it also produces
a non-vanishing disconnected four-point function which represents 
an additional source of noise for the bispectrum. 
Potentially this could severely hamper the measurement of $f_{\rm NL}$
and a detailed study of the effect is required.
Major improvements in the measurement of $f_{\rm NL}$ 
could be achieved by constructing optimal estimators 
that weigh more those triangular configurations that correspond to 
particularly high signal-to-noise ratios. 
We will address these issues in future work.

\acknowledgments
We thank Nicola Bartolo, Michele Liguori, Simon J. Lilly 
and Padelis Papadopoulos for helpful discussions.
AP thanks the organizers of the workshop on non-Gaussianity held on July 2006
in Trieste where the main results of this paper have been presented.

\appendix

\section{Perturbations in the gas temperature as a function of density fluctuations}
\label{PGT}

\begin{figure}
\plotone{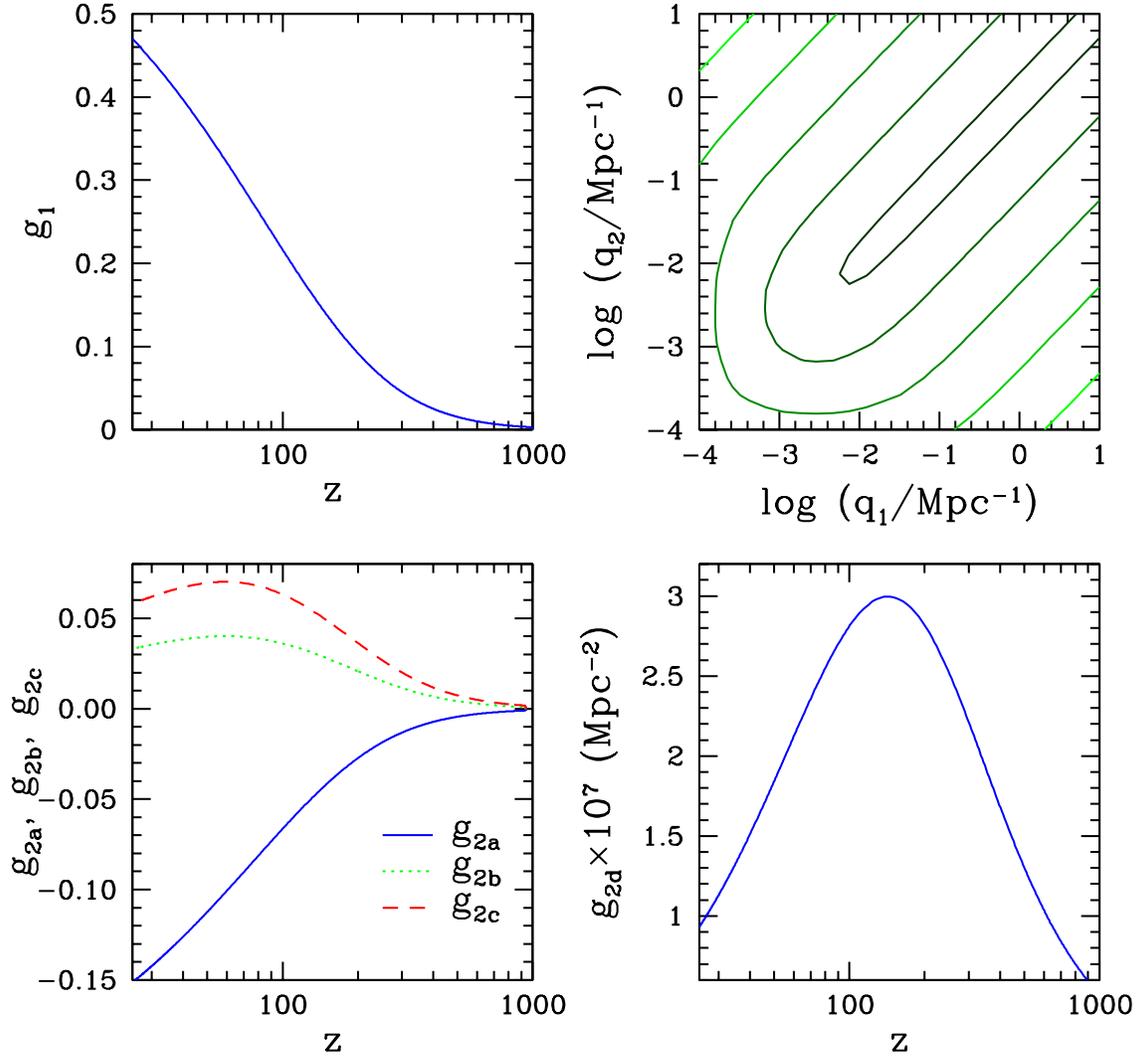}
\caption{\label{KERNELSFIG}
Redshift evolution and Fourier-space dependence of the kernels $\tilde{\GUNO}$ and $\tilde{\GDUE}$
relating temperature and density perturbations 
(see \S\ref{FLUCT} and Appendix \ref{PGT} for further details). 
Top left: redshift evolution of the first-order function $g_1(z)$.
Top right: contour levels of the kernel $\tilde{\GDUE}(\mathbf{q_1},\mathbf{q_2},z)$ at $z=50$ with 
$f_{\rm NL}=1$ and ${\mathbf q_1} \parallel {\mathbf q_2}$; consecutive levels increase by a factor of 10 
starting from 0.1 at the centre.
Bottom left: redshift evolution of the functions $g_{\rm 2a},g_{\rm 2b}$ and $g_{\rm 2c}$ 
introduced in equation (\ref{GDUESOLUTION}).
Bottom right: redshift evolution of the function $g_{\rm 2d}$ defined in equation (\ref{GDUESOLUTION}). 
}
\end{figure}

Equation \ref{evoTgas2} shows that
temperature and density perturbations in the cosmic distributions of baryons are related to each other. 
In this Appendix we want to express $\delta_T$ in terms of $\delta$ up to second order in perturbation
theory.
Our starting point is the functional expansion given in equation (\ref{GUNOGDUERS}) that we Fourier
transform, to obtain:
\be
\label{GUNOGDUEFS}
\tilde{\delta}_T(\mathbf{k},t)= \tilde{\GUNO}(\mathbf{k},t) \, \tilde{\delta}(\mathbf{k},t)+
\int d^3\mathbf{q_1} \int d^3\mathbf{q_2} \,\, \tilde{\GDUE}(\mathbf{q_1},\mathbf{q_2},t) \,\delta_{\rm{D}}(\mathbf{q_1}+\mathbf{q_2}-\mathbf{k})\, \tilde{\delta}(\mathbf{q_1},t) \, \tilde{\delta}(\mathbf{q_2},t)+...\;.
\ee
Performing a perturbative expansion up to second order gives:
\begin{eqnarray}
\tilde{\delta}_T^{(1)}(\mathbf{k},t)&=&\tilde{\GUNO}(\mathbf{k},t) \, \tilde{\delta}^{(1)}(\mathbf{k},t)\;,\\
\frac{1}{2}\,\tilde{\delta}_T^{(2)}(\mathbf{k},t)&=&\frac{1}{2}\,\tilde{\GUNO}(\mathbf{k},t) \, \tilde{\delta}^{(2)}(\mathbf{k},t)+
\int d^3\mathbf{q}\,\,\tilde{\GDUE}(\mathbf{q},\mathbf{k}-\mathbf{q},t) \,
\tilde{\delta}^{(1)}(\mathbf{q},t) \, \tilde{\delta}^{(1)}(\mathbf{k}-\mathbf{q},t)\;.
\end{eqnarray}
Thus, Taylor expanding equation (\ref{evoTgas2}) and
comparing terms of first perturbative order, one obtains:
\be
\label{GUNOEQ}
D_+(t)\,
\frac{\partial \tilde{\GUNO}(\mathbf{k},t)}{\partial t}
=-C(t) \, D_+(t)\, \tilde{\GUNO}(\mathbf{k},t) 
+\left(\frac{2}{3}-\tilde{\GUNO}(\mathbf{k},t) \right) \,\frac{\partial D_+(t)}{\partial t},
\ee
where we factorized the growth of linear overdensities as in equation (\ref{1BDFS}). 
Since this differential equation does not depend on $\mathbf{k}$ explicitly,
it follows that $\tilde{\GUNO}$ is only a function of time, i.e. $\tilde{\GUNO}(\mathbf{k},t)=g_1(t)$
and $\GUNO(\mathbf{x},t)=g_1(t)\,\delta_{\rm D}(\mathbf{x})$.
The corresponding function  $g_1(z)$ can be obtained performing a simple numerical integration 
and is plotted in Figure \ref{KERNELSFIG}.

The evolution equation for second-order fluctuations reads
\begin{eqnarray}
\label{GDUEEQ}
D^2_+(t)\,\frac{\partial \tilde{\GDUE}(\mathbf{q_1},\mathbf{q_2},t)}{\partial t}
&+&\left( 2 D_+(t)\frac{\partial D_+(t)}{\partial t} + C(t) D^2_+(t)\right)\tilde{\GDUE}(\mathbf{q_1},\mathbf{q_2},t)\\
\nonumber
&=& \left[\left(\frac{2}{3}- g_1 \right)\, 2 D_+(t)\frac{\partial D_+(t)}{\partial t} 
-\left(\frac{\partial g_1}{\partial t}+ g_1 C(t) \right)D^2_+(t)\right] \, \mathcal{K}(\mathbf{q_1},\mathbf{q_2},t)\\
\nonumber
&+&\left(\frac{2}{3}g_1-g_1 \right)2 D_+(t)\frac{\partial D_+(t)}{\partial t}
-\left(+\frac{\partial g_1}{\partial t}-g_1 C(t) \right),
\end{eqnarray}
where we used equations (\ref{2BDFS}), (\ref{GRAVITATIONAL}), (\ref{KERNEL}), and (\ref{PRIMORDIAL})
to write $\tilde\delta^{(2)}$ in terms of the kernel $\mathcal{K}(\mathbf{q_1},\mathbf{q_2},t)$.
The solution of this differential equation can be written as follows:
\begin{eqnarray}
\label{GDUESOLUTION}
\tilde{\GDUE}(\mathbf{q_1},\mathbf{q_2},t)&=&
g_{\rm 2a}(t)
+g_{\rm 2b}(t) \cos^2\theta_{12}
+g_{\rm 2c}(t)\left(\frac{q_1}{q_2}+\frac{q_2}{q_1} \right)\cos\theta_{12}\\
\nonumber
&+&f_{\rm{NL}}~g_{\rm 2d}(t)\frac{T(|{\bf q_1} + {\bf q_2}|)}{T(q_1)\,T(q_2)}\,\left( \frac{1}{q^2_1}+\frac{1}{q^2_2} + \frac{2}{q_1 q_2}\cos\theta_{12}  \right)\;,
\end{eqnarray}
where $\cos\theta_{12}=\hat{\mathbf q}_1\cdot\hat{\mathbf q}_2$.
The time-dependent terms $g_{\rm 2a}, g_{\rm 2b}, g_{\rm 2c}$ and $g_{\rm 2d}$ 
are plotted in Figure \ref{KERNELSFIG} as a function of redshift.

\section{Second order expansion of $\TBRIGHT$}
\label{SOET}
We present the most general expansion of $\TBRIGHT$ 
in terms of all the possible sources of spatial fluctuations 
up to second order.
The brightness temperature of the 21-cm background can be written as:

\begin{equation}
\TBRIGHT = \TBRIGHT^{(0)}+ \delta T^{(1)}_{\rm{21}} + 
\frac{1}{2}~\delta T^{(2)}_{\rm{21}} 
\end{equation}
where
\begin{eqnarray}
\label{21EXPANSION0}
\TBRIGHT^{(0)} & = &f_{\rm{0}}  \left( 1- \frac{\TCMB}{\mathcal{S}} \right) (1- \bar{x});\\
\delta T^{(1)}_{\rm{21}} & = &f^{\rm{b}}_1  \delta^{(1)} +f^{\rm{T}}_1  \delta^{(1)}_{\rm{T}}+ f^{x}_1  \delta^{(1)}_{x}+f^{\rm{v}}_1  \frac{\partial v^{(1)}}{\partial r}+f^{\rm{\alpha}}_1  \delta^{(1)}_{\rm{\alpha}};\\
\nonumber
\frac{1}{2}\delta T^{(2)}_{\rm{21}} & = &\frac{1}{2}f^{\rm{b}}_1  \delta^{(2)} +\frac{1}{2}f^{\rm{T}}_1  \delta^{(2)}_{\rm{T}} +\frac{1}{2} f^{x}_1  \delta^{(2)}_{x}+\frac{1}{2}f^{\rm{v}}_1  \frac{\partial v^{(2)}}{\partial r} + \frac{1}{2} f^{\rm{\alpha}}_1  \delta^{(2)}_{\rm{\alpha}} \\
\nonumber
& &+f^{\rm{bb}}_2  \delta^{(1)}\delta^{(1)} + f^{\rm{TT}}_2  \delta^{(1)}_{\rm{T}} \delta^{(1)}_{\rm{T}} +f^{xx}_2  \delta^{(1)}_{x}  \delta^{(1)}_{x}+f^{\rm{vv}}_2  \frac{\partial v^{(1)}}{\partial r}\frac{\partial v^{(1)}}{\partial r} + f^{\rm{\alpha}\rm{\alpha}}_2  \delta^{(1)}_{\rm{\alpha}}\delta^{(1)}_{\rm{\alpha}}+\\
& &+f^{{\rm b} x}_2  \delta^{(1)}\delta^{(1)}_{x} + f^{\rm{bT}}_2  \delta^{(1)}\delta^{(1)}_{\rm{T}}+f^{\rm{bv}}_2  \delta^{(1)} \frac{\partial v^{(1)}}{\partial r}+f^{\rm{b}\rm{\alpha}}_2  \delta^{(1)}\delta^{(1)}_{\rm{\alpha}}+ \\
\nonumber
& &+f^{{\rm {T}} x}_2 \delta^{(1)}_{\rm{T}} \delta^{(1)}_{x} +f^{\rm{Tv}}_2 \delta^{(1)}_{\rm{T}}  \frac{\partial v^{(1)}}{\partial r}+f^{\rm{T}\rm{\alpha}}_2  \delta^{(1)}_{\rm{T}}\delta^{(1)}_{\rm{\alpha}}+\\
\nonumber
&  & +f^{x\rm{v}}_2  \delta^{(1)}_{x} \frac{\partial v^{(1)}}{\partial r}+f^{x\rm{\alpha}}_2  \delta^{(1)}_{x}\delta^{(1)}_{\rm{\alpha}}+f^{\rm{v}\rm{\alpha}}_2  \frac{\partial v^{(1)}}{\partial r}\delta^{(1)}_{\rm{\alpha}}.
\end{eqnarray}
Here $\delta^{(n)}_i$ indicates the $n$-th order fluctuation in a given random field denoted by the subscript
$i$:
$\rm{b}$ stands for the baryon density, $\rm{T}$ for the gas temperature, $x$ for the hydrogen ionization fraction, $\rm{v}$ for the radial peculiar velocity, 
and $\alpha$ for the Ly$\alpha$ pumping efficiency. 

To simplify the notation, we denote zero-th order (i.e. unperturbed), time-dependent quantities with
overlined symbols and we indicate particular combinations of these zero-th order quantities
with a few special characters: 
\be
 \mathcal{C}= \phantom{a} \frac{4 ~\kappa_{\rm{10}}(\BTGAS) ~T_{*}}{3~ A_{\rm{10}} ~ \BTGAS}~  \bar{n}^0_{\rm HI}\;,
 \ \ \ \ \
 \mathcal{Y}_{\rm{C}}= \mathcal{C}(1+z)^3 (1-\bar{x})\;,
\ \ \ \ \ \
 \mathcal{Y}= \phantom{a}1+  \YA + \mathcal{Y}_{\rm{C}}\;,
\ee
\be
 \Delta \mathcal{T}=\phantom{a} \TCMB  - \BTGAS\;,
\ \ \ \ \ \ \ \
 \mathcal{S}=\phantom{a}\frac{1}{\mathcal{Y}} \left[ \TCMB +  \YA~ \BTGAS + \mathcal{Y}_{\rm{C}} \BTGAS  \right]\;.
\ee
We also introduce 
the symbols $\eta_1$ and $\eta_2$ to
parameterize the dependence of the collisional coupling rate $\kappa_{\rm{10}}$ 
(defined in Section \ref{spintempneuh}) up to second order on $\TGAS$:
\be
\kappa_{\rm{10}}=\kappa_{\rm{10}}(\BTGAS)~(\eta_1~\delta^{(1)}_{\rm{T}}+\frac{1}{2}~\eta_1 ~\delta^{(2)}_{\rm{T}}+\frac{1}{2}~\eta_2~\delta^{(1)}_{\rm{T}}~\delta^{(1)}_{\rm{T}}).
\ee
Adopting this notation,
the redshift-dependent weights which couple the brightness temperature to the various sources of fluctuations are: 

\begin{eqnarray}
&f_{\rm{0}}   &= \left( \frac{3 c^3 \hbar A_{\rm{10}}}{16 k_{\rm{B}} \nu_{\rm{21}}} \right) \frac{\bar{n}_{\rm HI}} {H_{\rm{0}}} {\left( \frac{1}{\Omega_{\rm{m}}} \right)}^{1/2} (1+z)^{1/2}\\
&f^{\rm{b}}_1  &=f_{\rm{0}}  \left[ \left( 1- \frac{\TCMB}{\mathcal{S}} \right) (1- \bar{x}) -\frac{\TCMB~\mathcal{C}~(1+z)^3 } {\mathcal{Y}^2 ~\mathcal{S}^2} \Delta \mathcal{T} (1- \bar{x})^2  \right] \\
&f^{\rm{T}}_1  &= f_{\rm{0}}  \left[ \frac{\TCMB~\mathcal{C}~(1+z)^3 } {\mathcal{Y} ~\mathcal{S}} \left( 1- 2~\frac{\Delta \mathcal{T}}{\mathcal{Y}~\mathcal{S}}\eta_1 \right)  (1- \bar{x})^2  + \frac{\TCMB}{\mathcal{Y} ~\mathcal{S}^2} (\YA~\BTGAS)(1- \bar{x}) \right] 
\end{eqnarray}
\begin{eqnarray}
&f^{x}_1  &=f_{\rm{0}}  \left[- \left( 1- \frac{\TCMB}{\mathcal{S}} \right) \bar{x} +\frac{\TCMB~\mathcal{C}~(1+z)^3 } {\mathcal{Y}^2 ~\mathcal{S}^2} \Delta \mathcal{T} (1- \bar{x}) \bar{x}  \right]\\
&f^{\rm{v}}_1  &=f_{\rm{0}}  \left[- \left( 1- \frac{\TCMB}{\mathcal{S}} \right) (1- \bar{x}) \frac{1+z}{H } \right]\\
&f^{\rm{\alpha}}_1  & =f_{\rm{0}}  \left[-\frac{\YA}{\mathcal{Y}^2 \mathcal{S}^2}\Delta \mathcal{T} \TCMB (1- \bar{x})\right]\\
&f^{\rm{bb}}_2 &=f_{\rm{0}}  \left[ -\frac{\TCMB~\mathcal{C}~(1+z)^3} {\mathcal{Y}^2 \mathcal{S}^2} \Delta \mathcal{T} (1- \bar{x})^2 + \frac{\TCMB~\mathcal{C}^2 ~(1+z)^6 }{\mathcal{S}^2 \mathcal{Y}^3} \Delta \mathcal{T} \left(1- \frac{\delta \mathcal{T}}{\mathcal{S} \mathcal{Y}} \right) (1- \bar{x})^3  \right] \\
&f^{\rm{TT}}_2  &=f_{\rm{0}}  \left[ \frac{\TCMB~\mathcal{C}~(1+z)^3 }{\mathcal{Y} ~\mathcal{S}}\left( 1-\eta_1+\frac{\Delta \mathcal{T}} {\mathcal{Y}~\mathcal{S}}~\eta_2 (1- \bar{x})^2 \right) + \right. \\
\nonumber
& &\left. \phantom{=f_{\rm{0}}  b} -\frac{\TCMB~\mathcal{C}~(1+z)^3}{\mathcal{Y}^2 \mathcal{S}^2} (\YA \BTGAS) \left( 1+\eta_1 \left( 1-4\frac{\Delta \mathcal{T}}{\mathcal{Y} ~\mathcal{S}}\right)  \right) (1- \bar{x})^2+ \right. \\ 
\nonumber
& &\left. \phantom{=f_{\rm{0}}  b}+\frac{\TCMB~\mathcal{C}^2 ~(1+z)^6 }{\mathcal{S} \mathcal{Y}^2} 
\left( -2 + 4~\left( \frac{\Delta \mathcal{T}}{\mathcal{S} \mathcal{Y}} \right)+ \left( 1-4 \left(\frac{\Delta \mathcal{T}} {\mathcal{S} \mathcal{Y}}\right)^2 \right)~\eta_1 \right) (1- \bar{x})^3+ \right.\\ 
\nonumber
& & \left. \phantom{=f_{\rm{0}}  b}-
\frac{\TCMB}{\mathcal{Y}^3 \mathcal{S}^2} (\YA~\BTGAS)^2(1- \bar{x})\right]  \\
&f^{xx}_2 &=f_{\rm{0}} \left[ +\frac{\TCMB~\mathcal{C}~(1+z)^3}{\mathcal{Y}^2 \mathcal{S}^2} \Delta \mathcal{T} {\bar{x}}^2 - \frac{\TCMB~\mathcal{C}^2 ~(1+z)^6 }{\mathcal{Y}^3~\mathcal{S}^2 } \Delta \mathcal{T} \left(1- \frac{\Delta \mathcal{T}}{\mathcal{Y}~\mathcal{S}} \right) (1- \bar{x})~{\bar{x}}^2 \right] \\
&f^{\rm{vv}}_2 &= f_{\rm{0}} \left[ \left( 1- \frac{\TCMB}{\mathcal{S}} \right) (1- \bar{x}) \frac{(1+z)^2}{H^2} \right] \\
&f^{\rm{\alpha}\rm{\alpha}}_2  & =f_{\rm{0}}  \left[+\frac{\YA}{\mathcal{Y}^3 \mathcal{S}^2}\Delta \mathcal{T}~ \TCMB \left( 1-\frac{\YA}{\mathcal{Y} ~\mathcal{S}}\Delta \mathcal{T}  \right) (1- \bar{x}) \right]\\
&f^{\rm{bT}}_2  & = f_{\rm{0}}  \left[ \frac{\TCMB~\mathcal{C}~(1+z)^3 } {\mathcal{Y} ~\mathcal{S}} \left(2-3~\left( \frac{\Delta \mathcal{T}}{\mathcal{Y} ~\mathcal{S}}\right)~\eta_1 \right) (1- \bar{x})^2+ \right. \\
\nonumber
& &\left. \phantom{=f_{\rm{0}}  b} -\frac{\TCMB~\mathcal{C}~(1+z)^3}{\mathcal{Y}^2 \mathcal{S}^2} (\YA \BTGAS)\left(1- 2\frac{\Delta \mathcal{T}}{\mathcal{S} \mathcal{Y}}\right)(1- \bar{x})^2+ \frac{\TCMB}{\mathcal{Y} ~\mathcal{S}^2} (\YA \BTGAS)(1- \bar{x}) +\right. \\
\nonumber
& &\left.\phantom{=f_{\rm{0}}  b}+\frac{ \TCMB\mathcal{C}^2(1+z)^6 }{\mathcal{Y}^2 \mathcal{S}} \left(-2+ \frac{\BTGAS}{\mathcal{S}}+2\frac{\Delta \mathcal{T}}{\mathcal{Y}~\mathcal{S}}(1+\eta_1)-4\left(\frac{\Delta \mathcal{T}}{\mathcal{Y}~\mathcal{S}} \right)^2 \eta_1 \right) (1- \bar{x})^3  \right] 
\end{eqnarray}
\begin{eqnarray}
&f^{{\rm b}x}_2 &=f_{\rm{0}}  \left[- \left( 1- \frac{\TCMB}{\mathcal{S}} \right) \bar{x} +3\frac{ \TCMB~\mathcal{C}~(1+z)^3 }{\mathcal{Y}^2 \mathcal{S}^2} \Delta \mathcal{T} (1- \bar{x})\bar{x}  
+ \right. \\ 
\nonumber
& &\left. \phantom{=f_{\rm{0}}  b} 
-2\frac{\TCMB~\mathcal{C}^2 ~(1+z)^6 }{ \mathcal{Y}^3\mathcal{S}^2} \Delta \mathcal{T} \left(1- \frac{\Delta \mathcal{T}}{\mathcal{Y}~\mathcal{S} } \right) (1- \bar{x})^2 {\bar{x}}  \right]\\
&f^{\rm{bv}}_2 &= f_{\rm{0}} \left[- \left( 1- \frac{\TCMB}{\mathcal{S}} \right) (1-\bar{x})  \frac{1+z}{H} +\frac{ \TCMB~\mathcal{C}~(1+z)^3 }{\mathcal{Y}^2 \mathcal{S}^2} \Delta \mathcal{T} (1- \bar{x})^2  \frac{1+z}{H}  \right]\\
&f^{\rm{b}\rm{\alpha}}_2  & =f_{\rm{0}} \left[-2 \frac{\TCMB~\mathcal{C}~(1+z)^3 }{\mathcal{Y}^2 \mathcal{S}^2}  \left(\YA\BTGAS+\frac{\YA}{\mathcal{Y}^2\mathcal{S}}\Delta \mathcal{T}^2 \right) (1-\bar{x})^2  -\frac{\YA}{\mathcal{Y}^2 \mathcal{S}^2}\Delta \mathcal{T}~\TCMB (1-\bar{x})\right]\\
&f^{{\rm T}x}_2 & =  f_{\rm{0}}  \left[ -\frac{\TCMB~\mathcal{C}~(1+z)^3 } {\mathcal{Y} ~\mathcal{S}} \left(2-3~\left( \frac{\Delta \mathcal{T}}{\mathcal{Y} ~\mathcal{S}}\right)~\eta_1 \right) (1- \bar{x}) \bar{x}+ \right. \\
\nonumber
& &\left. \phantom{=f_{\rm{0}}  b} +\frac{\TCMB~\mathcal{C}~(1+z)^3}{\mathcal{Y}^2 \mathcal{S}^2} (\YA \BTGAS)\left(1- 2\frac{\Delta \mathcal{T}}{\mathcal{S} \mathcal{Y}}\right)(1- \bar{x})\bar{x}- \frac{\TCMB}{\mathcal{Y} ~\mathcal{S}^2} (\YA \BTGAS)\bar{x} +\right. \\
\nonumber
& &\left.\phantom{=f_{\rm{0}}  b}-\frac{ \TCMB\mathcal{C}^2(1+z)^6 }{\mathcal{Y}^2 \mathcal{S}} \left(-2+ \frac{\BTGAS}{\mathcal{S}}+2\frac{\Delta \mathcal{T}}{\mathcal{Y}~\mathcal{S}}(1+\eta_1)-4\left(\frac{\Delta \mathcal{T}}{\mathcal{Y}~\mathcal{S}} \right)^2 \eta_1 \right) (1- \bar{x})^2 \bar{x} \right] \\
&f^{\rm{Tv}}_2 & =  f_{\rm{0}}  \left[ +2\frac{\TCMB~\mathcal{C}~(1+z)^3 }{\mathcal{Y}^2 \mathcal{S}^2} \Delta \mathcal{T} (1- \bar{x})~\eta_1\frac{1+z}{H} -\frac{ \TCMB~\mathcal{C}~(1+z)^3 } {\mathcal{Y}^2 \mathcal{S}^2} (\mathcal{Y} ~\mathcal{S})(1- \bar{x})^2\frac{1+z}{H} + \right. \\
\nonumber
& &\left. \phantom{=f_{\rm{0}}  b} -\frac{\YA}{\mathcal{Y} ~\mathcal{S}^2}\TCMB \BTGAS(1- \bar{x})\frac{1+z}{H}\right]
\end{eqnarray}
\begin{eqnarray}
&f^{T\rm{\alpha}}_2  & =f_{\rm{0}} \left[\frac{\TCMB~\mathcal{C}~(1+z)^3 }{\mathcal{Y}^2 \mathcal{S}^2}(\YA~\BTGAS)\left(1-2\eta_1  \right)(1-\bar{x})^2\eta_1 + \frac{\YA}{\mathcal{Y}^2 \mathcal{S}^2}
\BTGAS \TCMB\left( 1-\frac{\YA}{\mathcal{Y}}\right)+ \right.\\
\nonumber
& &\left. \phantom{=f_{\rm{0}}  b} +2\frac{\TCMB~\mathcal{C}~(1+z)^3 }{\mathcal{Y}~\mathcal{S}^2}\frac{\YA}{\mathcal{Y}^2} \Delta \mathcal{T}\left(1-2 \left(\frac{\Delta \mathcal{T}}{\mathcal{Y}~\mathcal{S}} \right)\eta_1  \right)(1-\bar{x})^2 + 2\frac{\YA}{\mathcal{Y}^3 \mathcal{S}^3} \BTGAS \Delta \mathcal{T}~\TCMB (1-\bar{x})\right]\\ 
&f^{x\rm{v}}_2 &=f_{\rm{0}}  \left[ \left( 1- \frac{\TCMB}{\mathcal{S}} \right) \bar{x}\frac{1+z}{H} -\frac{ \TCMB~\mathcal{C}~(1+z)^3 }{\mathcal{Y}^2 \mathcal{S}^2} \Delta \mathcal{T} (1- \bar{x})\bar{x}  \frac{1+z}{H}  \right].\\
& f^{x\rm{\alpha}}_2  & =f_{\rm{0}} \left[+2~ \frac{\TCMB~\mathcal{C}~(1+z)^3 }{\mathcal{Y}^2 \mathcal{S}^2}\left( (\YA~\BTGAS)+ \frac{\YA}{\mathcal{Y}^2\mathcal{S}}\Delta \mathcal{T}^2 \right)(1-\bar{x})\bar{x} +\frac{\YA}{\mathcal{Y}^2 \mathcal{S}^2}\Delta \mathcal{T}~ \TCMB\bar{x}\right]\\ 
&f^{\rm{v}\rm{\alpha}}_2  & =f_{\rm{0}} \left[\frac{\YA}{\mathcal{Y}^2 \mathcal{S}^2}\Delta \mathcal{T}~ \TCMB(1-\bar{x})~\frac{1+z}{H}\right]\;.
\end{eqnarray}

\clearpage
\end{document}